\documentclass[11pt]{article}
\usepackage{geometry}
\usepackage{enumerate}
\usepackage[backref, colorlinks,citecolor=blue,bookmarks=true]{hyperref}  
\usepackage{graphicx}
\usepackage{mathrsfs,amssymb,amsmath,textcomp,amsfonts,amsthm,bbm,fullpage}
\usepackage{color}
\usepackage{multirow,amsmath, dsfont, amscd, latexsym,color,amssymb,setspace}
\usepackage{ifthen}
\usepackage{enumerate}

\usepackage[usenames,dvipsnames]{xcolor}

\def\colorful{3}
\def\comments{0}

\ifnum\colorful=1

\newcommand{\red}[1]{{\color{red} {#1}}}
\newcommand{\green}[1]{{\color{green} {#1}}}

\newcommand{\newred}[1]{{\color{red} {#1}}}

\fi
\ifnum\colorful=0

\newcommand{\red}[1]{{{#1}}}
\newcommand{\green}[1]{{{#1}}}

\newcommand{\newred}[1]{{{#1}}}

\fi

\ifnum\colorful=2

\newcommand{\red}[1]{{{#1}}}
\newcommand{\newred}[1]{{\color{red} {#1}}}
\newcommand{\green}[1]{{{#1}}}

\fi

\ifnum\colorful=3

\newcommand{\red}[1]{{{#1}}}
\newcommand{\newred}[1]{{{#1}}}
\newcommand{\green}[1]{{{#1}}}

\newcommand{\newgreen}[1]{{\color{green} {#1}}}

\newcommand{\rnote}[1]{\footnote{{\bf \color{orange}Rocco:} {#1}}}
\newcommand{\anote}[1]{\footnote{{\bf \color{orange}Avishay:} {#1}}}
\newcommand{\pnote}[1]{\footnote{{\bf \color{orange}Parik:} {#1}}}
\fi

\newcommand{\poly}{\mathrm{poly}}
\newcommand{\eps}{\epsilon}

\newcommand{\acz}{\mathsf{AC^0}}
\newcommand{\dnf}{\mathsf{DNF}}
\newcommand{\cnf}{\mathsf{CNF}}

\newcommand{\ignore}[1]{}
\newcommand{\zo}{\{0,1\}}
\newcommand{\zon}{\{0,1\}^n}
\newcommand{\rgta}{\ensuremath{\rightarrow}}
\newcommand{\pmo}{\ensuremath \{ \pm 1\}}
\newcommand{\ff}[2]{\ensuremath \prod_{i=0}^{#2 -1}(#1 -i)}

\newcommand{\R}{\mathbb R}
\newcommand{\E}{\mathop{\mathbb{E}\/}}
\newcommand{\Ex}{\mathop{\mathbb{E}\/}}
\newcommand{\samp}{\ensuremath{\leftarrow}}
\newcommand{\calD}{{\cal D}}
\newcommand{\D}{\calD}
\newcommand{\calC}{{\mathcal{C}}}
\newcommand{\calS}{{\cal S}}
\newcommand{\calT}{{\cal T}}
\newcommand{\stot}{s_{\mathrm{tot}}}
\newcommand{\bX}{\mathbf{X}}
\newcommand{\bY}{\mathbf{Y}}

\newcommand{\sbkts}[1]{{\ensuremath \left[ #1 \right]}}
\newcommand{\bkts}[1]{{\ensuremath \left( #1 \right)}}

\newcommand{\x}{{\boldsymbol{x}}}
\newcommand{\bx}{\boldsymbol{x}}
\newcommand{\bolds}{\boldsymbol{s}} 
\newcommand{\bT}{\mathbf{T}}
\newcommand{\bd}{\boldsymbol{d}}
\newcommand{\Rkn}{\ensuremath {\mathcal{R}_{k,n}}}
\newcommand{\brho}{{\boldsymbol{\rho}}}
\newcommand{\la}{\langle}
\newcommand{\ra}{\rangle}
\newcommand{\eat}[1]{}
\newcommand{\mb}[1]{\ensuremath{\mathbf{#1}}}

\newcommand{\supp}{\mathrm{supp}}
\newcommand{\I}{{\mathbb{I}}}
\newcommand{\h}[1]{\ensuremath{\hat{#1}}}

\newcommand{\smax}{\ensuremath s}

\newcommand{\sk}{\ensuremath {s}^k}

\newcommand{\sffk}{\ensuremath s^{\underline{k}}}
\newcommand{\sff}[1]{\ensuremath s^{\underline{#1}}}

\DeclareMathOperator{\Ik}{\I^k}
\DeclareMathOperator{\Iffk}{\I^{\underline{k}}}

\DeclareMathOperator{\cdim}{cdim}
\DeclareMathOperator{\dt}{dt}
\newcommand{\bK}{\mathbf{K}}
\newcommand{\by}{\boldsymbol{y}}
\newcommand{\y}{{\boldsymbol{y}}}

\newcommand{\ind}[1]{\ensuremath \mathbf{1}(#1)}

\newcommand{\bart}{\overline{\ell(T)}}
\newcommand{\eqdef}{\stackrel{\textrm{def}}{=}}
\DeclareMathOperator{\ts}{ts}
\DeclareMathOperator{\s}{s}
\DeclareMathOperator{\wt}{wt}

\newcommand{\rta}{\ensuremath{\rightarrow}}

\newcommand{\W}{\mathbb{W}}
\newcommand{\B}{\{0,1\}}
\newcommand{\ent}{\mathbb{H}}
\newcommand{\Parity}{\text{\sc Parity}}
\newcommand{\Ham}{\text{\sc Ham}}

\newcommand{\OR}{\text{\sc Or}}

\newenvironment{Proof}{\medbreak
\noindent {\bf Proof:~}}{\unskip\nobreak\hfill\hskip 2em \qed\par\medbreak}

\newcounter{dummy} \numberwithin{dummy}{section}
\newtheorem{Thm}[dummy]{Theorem}
\newtheorem{Lem}[dummy]{Lemma}
\newtheorem{Cor}[dummy]{Corollary}

\newtheorem{Conj}[dummy]{Conjecture}
\newtheorem{Def}[dummy]{Definition}
\newtheorem{Ques}[dummy]{Question}

\newtheorem*{Conj*}{Conjecture}

\title{Degree and Sensitivity: tails of two distributions\footnote{The conference version of this paper will appear in CCC'2016 \cite{GSW16}.}}

\author{
Parikshit Gopalan\\
Microsoft Research\\
{\tt parik@microsoft.com} 
\and Rocco A. Servedio\thanks{Supported by NSF grants CCF-1319788 and CCF-1420349.}\\
Columbia University\\
{\tt rocco@cs.columbia.edu}
\and Avishay Tal\thanks{
Supported by the Simons Foundation and by NSF grant CCF-1412958.}
\\
IAS, Princeton\\
{\tt avishay.tal@gmail.com}\\
\and  Avi Wigderson\thanks{This research was partially supported by NSF grant CCF-1412958.}
\\
IAS, Princeton\\
{\tt avi@ias.edu}
}

\begin{document}

\maketitle

 \thispagestyle{empty}

\begin{abstract}

The \emph{sensitivity} of a Boolean function $f$ is the maximum, 
\green{ over all inputs $x$, of the number of sensitive coordinates of $x$ (namely the number of Hamming neighbors of $x$ with different $f$-value).}  The well-known {\em sensitivity conjecture} of Nisan (see also Nisan and Szegedy) states that every sensitivity-$s$ Boolean function can be computed by a  polynomial over the reals of degree $\poly(s).$   The best known upper bounds on degree, however, are exponential rather than polynomial in $s$.

Our main result is an approximate version of the conjecture: every  Boolean function with sensitivity $s$ can be $\eps$-approximated (in $\ell_2$) by a polynomial whose degree is $O(s \cdot \log(1/\eps))$. This is the first improvement on the folklore bound of $s/\eps$. Further, we show that improving the bound to $O(s^c \cdot \log(1/\eps)^\gamma)$ for any $\gamma < 1$ and any $c > 0$ will imply the sensitivity conjecture. Thus our result is essentially the best one can hope for without proving the conjecture.\ignore{\anote{Added}} We apply our approximation result to obtain a new learning algorithm for insensitive functions, as well as new bounds on the Fourier coefficients and the entropy-influence conjecture for them.

\ignore{\anote{Shortened and added applications}}
\ifnum\comments=1
\rnote{Our abstract is very long, and this paragraph is all about techniques; shall we remove?}
\fi
We prove our result via a new ``switching lemma for low-sensitivity functions'' which establishes that a random restriction of a low-sensitivity function is very likely to have low decision tree depth. This is analogous to the well-known switching lemma 
\green{for $\acz$ circuits.} Our proof analyzes the combinatorial structure of the graph $G_f$ of sensitive edges of a Boolean function $f$.
We introduce new parameters of this graph such as {\em tree sensitivity}, study their relationship, and use them to show that the graph of a function  
\green{ of full degree} must be sufficiently complex, and that random restrictions of low-sensitivity functions are unlikely to yield complex graphs. 

We postulate a robust analogue of the sensitivity conjecture: if {\bf most} inputs to a Boolean function $f$ have low sensitivity, then {\bf most} of the Fourier mass of $f$ is concentrated on small subsets. We prove a lower bound on tree sensitivity in terms of decision tree depth, and show that a polynomial strengthening of this lower bound implies the robust conjecture.\ignore{  show that it is implied by a strengthening of the lower bound for tree sensitivity in terms of decision tree depth.} We feel that studying the graph $G_f$ is interesting in its own right, and we hope that some of the notions and techniques we introduce in this work will be of use in its further study.

\end{abstract}
\newpage

\setcounter{page}{1}



\section{Introduction} \label{sec:intro}

The smoothness of a continuous function captures how gradually it
changes locally (according to the metric of the underlying space). For
Boolean functions on $\zon$, a natural analog is {\em
  sensitivity}, capturing how many neighbors of a point have different
function values. More formally, the \emph{sensitivity} of $f:\zon \rgta \pmo$ at input $x \in \zon$, written $s(f,x)$,
is   the number of neighbors $y$ of $x$ in the Hamming cube $\zon$ such that $f(y) \neq f(x).$
The \emph{max sensitivity} of $f$, written $s(f)$ and often referred to simply as the ``sensitivity of $f$'', is defined as
$s(f)  = \max_{x\in \zon}s(f,x).$
Hence we have $0 \leq s(f) \leq n$ for every $f:\zon \rgta \pmo$; while not crucial, it may be helpful to consider this parameter as ``low'' when e.g. either
$s(f) \leq (\log n)^{O(1)}$ or $s(f)\leq n^{o(1)}$ (note that both these notions of ``low'' are robust up to polynomial factors).

A well known conjecture, sometimes referred to as the ``sensitivity conjecture,''  states that every smooth Boolean function is
computed by a low degree real polynomial, specifically of degree polynomial in its sensitivity.  This conjecture was first posed  in the form of  a question by Nisan \cite{Nisan:91} and Nisan and Szegedy  \cite{NisanSzegedy:94} but is now (we feel) widely believed to be true:

\begin{Conj}\cite{Nisan:91, NisanSzegedy:94}
\label{conj:N}
There exists a constant $c$ such that every Boolean function $f$ is computed by a polynomial of degree $\deg(f) \leq s(f)^c$. 
\end{Conj}

Despite significant effort (\cite{KK, AmbainisBGMSZ:14,AmbainisP:14,AmbainisPV:15,AmbainisV:15}) the best upper bound on degree in terms of sensitivity is exponential.
Recently several consequences of Conjecture \ref{conj:N}, e.g. that every $f$ has a \emph{formula} of depth at most $\poly(s(f))$, have been unconditionally established in \cite{GNSTW}. Nisan and Szegedy proved the converse, that every Boolean function satisfies $s(f) = O(\deg(f)^2)$. 

In this work, we make progress on Conjecture \ref{conj:N} by showing that functions with low max sensitivity are very well \emph{approximated} (in $\ell_2$) by low-degree polynomials. We exponentially improve the folklore $O(s/\eps)$ degree bound (which follows from average sensitivity and Markov's inequality) by replacing the $1/\eps$ error dependence with $\log(1/\eps)$. The following is our main result:

\begin{Thm}
\label{thm:main}
For any Boolean function $f: \zo^n \to \pmo$ and any $\eps > 0$, there exists a polynomial $p:\zo^n \rgta \R$ with $\deg(p) \leq O(s(f)\cdot \log(1/\eps))$ such that
$\E_{x \in  \zo^n}[|p(x) - f(x)|^2] \leq \eps.$\footnote{The conference version of this paper \cite{GSW16} proves a weaker bound of $O(s(f)(\log(1/\eps)^3))$.}
\end{Thm}

One might wonder if the dependence  on $\eps$
can be improved further.
We observe that a bound of $O(s^c \cdot \log(1/\eps)^\gamma)$ for any constants $\gamma < 1$ and $c > 0$ implies Conjecture \ref{conj:N}. Thus Theorem \ref{thm:main} gets essentially the best bound one can hope for without proving Conjecture \ref{conj:N}.
Furthermore, we show that a bound of  $O(s \cdot \log(1/\eps)^{\gamma})$ for a constant $\gamma<1$ does not hold for a family of Boolean functions based on the Hamming code. Hence, for $c=1$, Theorem~\ref{thm:main} is essentially tight.

En route to proving this result, we make two related contributions which we believe are interesting in themselves:
\begin{itemize}
\item Formulating a robust variant of the sensitivity conjecture (proving which would generalize Theorem \ref{thm:main}).
\item Defining and analyzing some natural graph-theoretic complexity measures, essential to our proof and which we believe may hold the key to progress on the original and robust sensitivity conjectures.
\end{itemize}

\subsection{A robust variant of the sensitivity conjecture} \label{sec:sdfd}

A remarkable series of developments, starting with \cite{Nisan:91},
showed that real polynomial degree is an extremely versatile complexity
measure:  it is polynomially related to many other 
complexity measures for Boolean functions, including PRAM complexity,
block sensitivity, certificate complexity, deterministic/randomized/quantum decision tree
depth, and
approximating polynomial degree (see \cite{BuhrmandeWolf:02,Chicago} for details on many of these relationships). 
Arguably the one natural complexity
measure that has defied inclusion in this equivalence class is
sensitivity.  Thus, there are many equivalent
formulations of Conjecture \ref{conj:N}; indeed, Nisan's original formulation was
in terms of sensitivity versus block sensitivity \cite{Nisan:91}.  

Even though progress on it has been slow, over the years Conjecture \ref{conj:N} has become a well-known open question in the study of Boolean functions. It is natural to ask \emph{why} this is an important  question: will a better understanding of sensitivity lead to new insights into Boolean functions that have eluded us so far? Is sensitivity qualitatively different from the other concrete complexity measures that we already understand?

We believe that the answer is yes, and in this paper we make the case that Conjecture \ref{conj:N} is just the (extremal) tip of the iceberg: it hints at deep connections between the \emph{combinatorial} structure of a Boolean function $f$, as captured by the graph $G_f$ of its sensitive edges in the hypercube, and the \emph{analytic} structure, as captured by its Fourier expansion. This connection is already the subject of some of the key results in the analysis of Boolean functions, such as \cite{KKL:88,Friedgut:98}, \green{as well as important open problems like the ``entropy-influence'' conjecture~\cite{FrKa:96} and its many consequences. }

Given any Boolean function $f$, we conjecture a connection between the distribution of the sensitivity of a random vertex in $\zo^n$ and the distribution of $f$'s Fourier mass.  This conjecture, which is an important motivation for the study in this paper, is stated informally below:

\medskip

\noindent {\bf Robust Sensitivity Conjecture (Informal Statement):} {\em If {\bf most} inputs to a Boolean function $f$ have low sensitivity, then {\bf most} of the Fourier mass of $f$ is concentrated on small subsets. }

\medskip

Replacing both occurrences of {\bf most} by {\em all} we recover Conjecture \ref{conj:N}, and hence the statement may be viewed as a robust formulation of the sensitivity conjecture.  Theorem \ref{thm:main} corresponds to replacing the first {\bf most} by {\em all}.  There are natural classes of functions 
which do not have low max sensitivity, but for which most vertices have low sensitivity; the robust sensitivity conjecture is relevant to these functions while the original sensitivity conjecture is not.
(A prominent example of such a class is $\acz$, for which the results of~\cite{LMN:93} establish a weak version of the assumption (that most inputs have low sensitivity) and the results of \cite{LMN:93,Tal:14tightbounds} establish a strong version of the conclusion (Fourier concentration).)

In order to formulate a precise statement, for a given Boolean function $f: \zon \to \pmo$ we consider the random experiment which samples from the following two distributions:

\begin{enumerate}
\item The Sensitivity distribution: sample a uniform random vertex $\bx \in \zo^n$ and let $\bolds  = s(f,\bx)$.
\item The Fourier distribution: sample a subset $\bT \subset [n]$ with probability $\hat{f}(\bT)^2$ and let $\bd = |\bT|$.
\end{enumerate}
We conjecture a close relation between the $k^{th}$ moments of these random variables:
\begin{Conj}[Robust Sensitivity Conjecture]
\label{conj:moments}
For all Boolean functions $f$ and all integers $k \geq 1$, there is a constant $a_k$ such that $\E[\bd^k] \leq a_k\E[\bolds^k]$. 
\end{Conj}
The key here is that there is no dependence on $n$.  To see the connection with the informal statement above, if a function has low sensitivity for most $x \in \zo^n$, then it must have bounded $k^{th}$ sensitivity moments for fairly large $k$; in such a case, Conjecture \ref{conj:moments} implies a strong Fourier concentration bound by Markov's inequality. The classical Fourier expansion for average sensitivity tells us that when $k=1$, $\E[\bolds] = \E[\bd]$. It is also known that $\E[\bolds^2] = \E[\bd^2]$ (see e.g. \cite[Lemma 3.5]{CKLS15}), but equality does not hold for $k \geq 3$. Conjecture \ref{conj:moments} states that if we allow constant factors depending on $k$, then one direction still holds. 

\green{It is clear that Conjecture~\ref{conj:moments} (with $a_k$ a not-too-rapidly-growing function of $k$) is a strengthening of our Theorem~\ref{thm:main}. To see its relation to  Conjecture~\ref{conj:N} observe that Conjecture \ref{conj:N} implies that for $k \to \infty$, $\E[\bd^k] \leq a^k(\E[\bolds^k])^{b}$ for constants $a, b$. On the other hand, via Markov's inequality, Conjecture \ref{conj:moments} only guarantees Fourier concentration rather than small degree for functions with small sensitivity. Thus the robust version Conjecture~\ref{conj:moments} seems incomparable to Conjecture \ref{conj:N}.}

It is possible that the reverse direction of the robust conjecture also holds: for every $k$ there exists $a'_k$ such that $\E[\bolds^k] \leq a'_k\E[\bd^k]$; settling this is an intriguing open question.  \ignore{
We note that the Nisan-Szegedy result that $s(f) \leq O(\deg(f)^2)$ implies that as $k\to \infty$ we have $\E[\bolds^k] \leq C^k \E[\bd^k]^2$ for some constant $C.$\anote{When you say $k\to \infty$, do you mean that $n$ is fixed and $k$ goes to infinity?
If $k$ goes to infinity and $n$ goes faster to infinity I don't see why the inequality holds.}
}

\green{
Both our proof of Theorem \ref{thm:main}, and our attempts at Conjecture \ref{conj:moments}, follow the same general path.
We apply random restrictions, which reduces these statements to analyzing some natural new graph-theoretic complexity measures  of Boolean functions. These measures are relaxations of sensitivity: they look for occurrences of various subgraphs in the sensitivity graph, rather than just high degree vertices. We establish (and conjecture) connections between different graph-theoretic measures and decision tree depth (see Theorem \ref{thm:dt-gives-pw}, which relates decision tree depth and the length of ``proper walks'', and Conjecture \ref{conj:1}, which conjectures a relation between ``tree sensitivity'' and decision tree depth). These connections respectively enable the proof of Theorem \ref{thm:main} and provide a simple sufficient condition implying Conjecture \ref{conj:moments}, which suffices to prove the conjecture for $k =3$ and $4$. We elaborate on this in the next subsection.
We believe that these new complexity measures are interesting and important in their own right, and that understanding them better may lead to progress on Conjecture \ref{conj:N}.
}

\subsection{Random restrictions and graph-theoretic complexity measures}
\label{sec:graph}

\green{In this subsection we give a high level description of our new complexity measures and perspectives on the sensitivity graph and of how we use them to approach Conjecture \ref{conj:moments} and prove Theorem \ref{thm:main}.} As both have the same conclusion, namely strong Fourier concentration, we describe both approaches together until they diverge. This leads to analyzing two different graph parameters (as we shall see, the stronger assumption of Theorem~\ref{thm:main} allows the use of a weaker graph parameter that we can better control).

First we give a precise definition of the {\em sensitivity graph}:  to every Boolean function $f$ we associate a graph $G_f$
whose vertex set is $\zo^n$ and whose edge set  $E$ consists of all
edges $(x,y)$ of the hypercube that have $f(x) \neq f(y)$.%
\ignore{
\anote{Should we consider adding this as a footnote? An alternative way to define $G_f$ is to consider $g(x) = f(x) \oplus \Parity(x_1, \ldots, x_n)$. Since $\Parity$ gets different values on any two neighbors of the Hypercube, $x$ is a neighbor of $y$ in $G_f$ iff $d(x,y)=1$ and $g(x) = g(y)$.
Let $V_0 = \{x\in \{0,1\}^n :g(x)=0\}$ and $V_1 = \{x\in \{0,1\}^n :g(x)=1\}$. Then, $G_f$ is the union of subgraphs of the Boolean hypercube induced by $V_0$ and $V_1$.
}
}
Each edge is labelled by the coordinate in $[n]$ at which $x$ and $y$ differ. The degree of
vertex $x$ is exactly $s(f,x)$, and  the maximum degree of $G_f$ is 
$s(f)$.

The starting point of our approach 
is to reinterpret the moments of the degree and sensitivity distributions of $f$ in terms of its random restrictions. Let $\Rkn$  denote the distribution over random restrictions that leave exactly $k$ of the $n$ variables unset and set the rest uniformly at random. We first show, in Section \ref{sec:moments}, that the $k^{th}$ moment of the sensitivity distribution controls the probability that a random restriction $f_\brho$ of $f$, where $\brho \leftarrow \Rkn,$ has full sensitivity (Theorem \ref{thm:sk}). Similarly, moments of the Fourier distribution capture the event that $f_\brho$ has full degree (Theorem \ref{thm:ik}). (We note that Tal has proved a result of a similar flavor; \cite[Theorem 3.2]{Tal14quantum} states that strong Fourier concentration of $f$ implies that random restrictions of $f$ are unlikely to have high degree.)

\subsubsection*{Random restrictions under sensitivity moment bounds:  Towards Conjecture~\ref{conj:moments}}

Given Theorems \ref{thm:sk} and \ref{thm:ik}, Conjecture \ref{conj:moments} may be rephrased as saying that if a function $f$ has low sensitivity moments, then a random restriction $f_\brho$ is unlikely to have full degree. Some intuition supporting this statement is that the sensitivity graphs of functions with full degree should be ``complex'' (under some suitable complexity measure), whereas the graph of $f_\brho$ is unlikely to be ``complex'' if $f$ has low sensitivity moments. \green{More precisely, the fact that $G_f$ has no (or few) vertices of high degree suggests that structures with many sensitive edges in distinct directions will \red{not} survive a random restriction.}

Some evidence supporting this intuition is given by Theorem \ref{thm:sk}, which tells us that if $f$ has low sensitivity moments then  $f_\brho$ is unlikely to have full sensitivity. If full degree implied full sensitivity then we would be done, but this is false as witnessed e.g. by the three-variable majority function and by composed variants of it.  (Conjecture \ref{conj:N} asserts that the gap between degree and sensitivity is at most polynomial, but of course we do not want to invoke the conjecture!) This leads us in Section \ref{sec:t} to consider our first relaxation of sensitivity, which we call \emph{tree-sensitivity}.  To motivate this notion, note that a vertex with sensitivity $k$ is simply a star with $k$ edges in the sensitivity graph. We relax the star requirement and consider all {\em sensitive trees}: trees of sensitive edges (i.e. edges in $G_f$) where every edge belongs to a \emph{distinct} coordinate direction (as is the case, of course, for a star).  \red{Analogous to the usual notion of sensitivity, the tree sensitivity of $f$ at $x$ is the size of the largest sensitive tree containing $x$, and the tree sensitivity of $f$ is the maximum tree sensitivity of $f$ at any vertex.}

Theorem \ref{thm:ts} shows that the sensitivity moments of $f$ control the probability that $f_\rho$ has full tree sensitivity.  Its proof crucially uses  a result by Sidorenko \cite{Sidorenko:94} on counting homomorphisms to trees. \green{Theorem \ref{thm:ts} would immediately imply Conjecture \ref{conj:moments} if every function of degree $k$ must have tree sensitivity $k$.  \newred{(This is easily verified for $k=3,4$, which, as alluded to in the previous subsection, gives Conjecture \ref{conj:moments} for those values of $k$.)}
The best we can prove, though, is a tree sensitivity lower bound of $\Omega(\sqrt{k})$ (Theorem \ref{thm:stree-dtree})}; \red{the proof of this lower bound uses notions of maximality and ``shifting'' of sensitive trees that we believe may find further application in the study of tree sensitivity.} We conjecture that full degree does imply full tree sensitivity, implying Conjecture \ref{conj:moments}. This is a rare example where having a precise  bound between the two complexity measures (rather than a polynomial relationship) seems to be important.

\subsubsection*{Random restrictions under max sensitivity bounds:  Proving Theorem~\ref{thm:main}}

Next, we aim to prove \emph{unconditional} moment bounds on the Fourier distribution of low sensitivity functions \green{and thereby obtain Theorem~\ref{thm:main}}. \red{Towards this goal, in Section \ref{sec:pw} we relax the notion of  tree sensitivity and study certain walks in the Boolean hypercube that we call \emph{proper walks}:   these are walks such that every time a coordinate direction is explored for the first time, it is along a sensitive edge.}
We show in Theorem \ref{thm:dt-gives-pw} that having full \red{decision tree depth} implies the existence of a very short \red{(length $O(n)$)} proper walk containing sensitive edges along every coordinate. \green{In Lemma \ref{lem:switching encoding}, we analyze random restrictions to show that such a structure is unlikely to survive in the remaining subcube of unrestricted variables.  This may be viewed 
as  a ``switching lemma for low-sensitivity functions'', which again may be independently interesting (note that strictly speaking this result is not about switching from a $\dnf$ to a $\cnf$ or vice versa, but rather it upper bounds the probability that a restricted function has large decision tree depth, in the spirit of standard ``switching lemmas'').
It yields Theorem \ref{thm:main} via a rather straightforward argument. The analysis requires an upper bound on the maximum sensitivity because we do not know an analogue of Sidorenko's theorem for proper walks. 
}

\subsection{Some high-level perspective}

An important goal of this work is to motivate a better understanding of the combinatorial structure of the sensitivity graph $G_f$ associated with a Boolean function. In our proofs other notions suggest themselves beyond tree sensitivity and proper walks, most notably the {\em component dimension} of the graph, which may be viewed as a further relaxation of sensitivity. Better relating these measures to decision tree depths, as well as to each other, remains intriguing, and in our view promising, for making progress on Conjecture~\ref{conj:N} and Conjecture~\ref{conj:moments} and for better understanding Boolean functions in general. We hope that some of the notions and techniques we introduce in this work will be of use to this goal.

\green{Another high level perspective relates to ``switching lemmas''. As mentioned above, we prove here a new result of this kind, showing that under random restrictions low sensitivity functions have low decision tree depth with high probability. The classical switching lemma shows the same for small width $\dnf$ (or $\cnf$) formulas (and hence for $\acz$ circuits as well). Our proof is quite different than the standard proofs, as it is essentially based on the combinatorial parameters of the sensitivity graph. Let us relate the assumptions of both switching lemmas.  On the one hand, by the sensitivity Conjecture~\ref{conj:N} (which we can't use, and want to prove), low sensitivity should imply low degree and hence low decision tree depth and small $\dnf$ width. On the other hand, small $\dnf$ width (indeed small, shallow circuits) imply (by \cite{LMN:93}) low {\em average} sensitivity, which is roughly the assumption of the robust sensitivity Conjecture~\ref{conj:moments}. As it turns out, we can use our combinatorial proof of our switching lemma to derive a somewhat weaker form of the original switching lemma, and also show that the same combinatorial assumption (relating tree sensitivity to decision tree depth) which implies Conjecture~\ref{conj:moments} would yield a nearly tight form of the original switching lemma. This lends further motivation to the study of these graph parameters.
}

Another conjecture formalizing the maxim that  {\em low sensitivity implies Fourier concentration} is the celebrated Entropy-Influence conjecture of Freidgut and Kalai \cite{FrKa:96} which posits the existence of a constant $C$ such that
$\ent(\bT) \leq C\E[\bolds]$
where $\ent(.)$ denotes the entropy function of a random variable.\footnote{Recall that the entropy $\ent(\bT)$ of the random variable $\bT$ is 
\[
\ent(\bT) = \sum_{T \subseteq [n]} \Pr[\bT=T] \log_2 {\frac 1 {\Pr[\bT=T]}}.
\]
} The conjecture states that functions with low sensitivity on average (measured by $\E[\bolds] = \E[\bd]$) have their Fourier spectrum concentrated on a few coefficients, so that the entropy of the Fourier distribution is low. However, unlike in Conjecture \ref{conj:moments} the degree of those coefficients does not enter the picture.

\medskip

\paragraph*{Organization.} We present some standard preliminaries and notation in Section \ref{sec:prelims}. Section \ref{sec:moments} proves Theorems \ref{thm:sk} and \ref{thm:ik} which show that degree and sensitivity moments govern the degree and sensitivity respectively of random restrictions. 

In Section \ref{sec:t} we study tree sensitivity. Section \ref{sec:ts-dt} relates it to other complexity measures, while Section \ref{sec:ts} relates tree sensitivity of a random restriction to the sensitivity moments. Section \ref{sec:pw} introduces proper walks and uses them to show Fourier concentration for low sensitivity functions.  We define proper walks in Section \ref{sec:def-pw}, and use them to analyze random restrictions of low-sensitivity functions in section \ref{sec:pw-rr}. We prove Theorem \ref{thm:main}  in \ref{sec:fourier-conc} and analyze its tightness in \ref{sec:tightness}. Section \ref{sec:pw} uses results from Section \ref{sec:ts-dt} but is independent of the rest of Section \ref{sec:t}. 

We derive applications to learning  in Section~\ref{sec:learn}, to the Entropy-Influence conjecture in Section~\ref{sec:Fourier-Entropy}, and present an approach to proving the Switching lemma for $\dnf$s via sensitivity moments in Section \ref{sec:sl-moments}. In Section~\ref{sec:ex} we present a family of functions that demonstrates the tightness of Theorem~\ref{thm:main}, and some additional examples, and highlight some open problems in Section \ref{sec:open}.




\section{Preliminaries}
\label {sec:prelims}

\noindent{\bf The Fourier distribution.} Let $f: \zo^n \rightarrow \pmo$ be a Boolean function. We define the usual
inner product on the space of such functions by
$\la f,g \ra = \Ex_{\x \samp \zo^n}[f(\x)g(\x)]$.
For $S \subseteq [n]$ the parity function $\chi_S$ is $\chi_S(x) = (-1)^{\sum_{i \in S}x_i}$.
The Fourier expansion of $f$ is given by
$f(x) = \sum_{S \subset [n]} \hat{f}(S) \chi_S(x),$ where
$\hat{f}(S) = \la f, \chi_S \ra.
$
By Parseval's identity we have $\sum_{S \subseteq [n]}\hat{f}(S)^2 =
1$. This allows us to view any Boolean function $f$ as inducing a probability
distribution $\D_f$ on subsets $S \subseteq [n]$, given by
$\Pr_{\mb{R} \samp \D_f}[\mb{R} = S] = \hat{f}(S)^2.$
We refer to this as the \emph{Fourier distribution}. We define $\supp(f)
\subseteq 2^{[n]}$ as $\supp(f) = \{S \subseteq [n]: \hat{f}(S)^2 \neq 0\}$.
The Fourier expansion of $f$ can be viewed as expressing $S$ as a
multilinear polynomial in $x_1,\ldots,x_n$, so that $\deg(f) = \max_{S \in \supp(f)}|S|$.
Viewing $\D_f$ as a probability distribution on $2^{[n]}$, we
define the following quantities which we refer to as ``influence moments''
of $f$:

\begin{align}
\label{eq:Ik-def}
\Ik[f] & = \Ex_{\mb{R} \samp \D_f}\left[{|\mb{R}|^k}\right] =  \sum_{S}\h{f}(S)^2{|S|^k}, \\
\label{eq:Iffk-def}
\Iffk[f] & = \Ex_{\mb{R} \samp \D_f} 
\left[ \ff{|\mb{R}|}{k}\right] =  \sum_{|S| \geq  k}\h{f}(S)^2\ff{|S|}{k}.
\end{align}

We write $\deg_\eps(f)$ to denote the minimum $k$ such that
$\sum_{S \subseteq [n]; |S| \geq k}\h{f}(S)^2 \leq \eps.$
It is well known that $\deg_\eps(f) \leq k$ implies the existence of a
degree $k$ polynomial $g$ such that $\Ex_{\x}[(f(\x) - g(\x))^2] \leq
\eps$; $g$ is obtained by truncating the Fourier expansion of $f$ to
level $k$.

\medskip

\noindent{\bf The sensitivity distribution.} We use $d(\cdot,\cdot)$ to denote Hamming distance on $\zo^n$.
The $n$-dimensional hypercube $H_n$ is the graph with vertex set
$V = \zo^n$ and $\{x,y\} \in E$ if $d(x,y) =1$. For $x \in \zo^n$, let
$N(x)$ denote its neighborhood in $H_n$. As described in Section \ref{sec:intro}, the \emph{sensitivity} of a function
$f$ at point $x$ is defined as
$s(f,x) = |\{y \in N(x): f(x) \neq f(y)\}|$, and
the (worst-case) sensitivity of $f$, denoted $s(f)$, is defined as
$s(f) = \max_{x \in \zo^n}s(f,x).$
Analogous to (\ref{eq:Ik-def}) and (\ref{eq:Iffk-def}), we define the quantities $s^k(f)$ and $\sffk(f)$ which we refer to as
``sensitivity moments'' of $f$:
\begin{align}
\label{eq:sk-def}
\sk(f) & = \Ex_{\x \samp \zo^n }\left[ s(f,\x)^k\right], \quad \quad \quad
\sffk(f) = \Ex_{\x \samp \zo^n} \left[ \ff{s(f,\x)}{k}\right].
\end{align}

With this notation, we can restate Conjecture \ref{conj:moments} (with a small modification) as 
\begin{Conj*}
(Conjecture \ref{conj:moments} restated)
For every $k$, there exists constants $a_k, b_k$ such that
$ \I^k(f) \leq a_k\sk(f) +b_k.$
\end{Conj*}
The reason for the additive constant $b_k$ is that for all non-negative integers $x$, we have
$$\ff{x}{k} \leq x^k \leq e^k\ff{x}{k} + k^k.$$ 
Hence allowing the additive factor lets us freely interchange $\I^k$ with $\Iffk$ and $\sk$ with $\sffk$ in the statement of the Conjecture.
We note that $\I^1[f] = \I^{\underline{1}}[f] = s^1(f) = s^{\underline{1}}(f)$, and as stated earlier it is not difficult to show that $\I^2[f] = s^2(f)$ (see e.g. Lemma 3.5 of \cite{CKLS15}).  However, in general $\I^k(f) \neq s^k(f)$ for $k \geq 3$ (as witnessed, for example, by the AND function).

\medskip

\noindent {\bf Some other complexity measures.}
We define $\dim(f)$ to be the number of variables that $f$
depends on and $\dt(f)$ to be the smallest depth of a deterministic decision tree computing $f$. 
In particular $f:\zo^n \rgta \pmo$ has $\dim(f) =n$ iff
$f$ is sensitive to every co-ordinate, and has $\dt(f)=n$ iff $f$ is evasive. It is easy to see that $\deg(f)
\leq \dt(f) \leq \dim(f)$ and $s(f) \leq \dt(f).$




\section{Random restrictions and moments of degree and sensitivity} \label{sec:moments}

We write $\Rkn$ to denote the set of all restrictions that leave
exactly $k$ variables live (unset) out of $n$.  
A restriction $\rho
\in \Rkn$ is viewed as a string in $\{0,1,\star\}^n$ where $\rho_i =
\star$ for exactly the $k$ live variables. We denote the set of live variables by $L(\rho)$, and 
we use $f_\rho:\zo^{L(\rho)} \rgta \pmo$ to denote the resulting
restricted function. We use $C(\rho) \subseteq \zo^n$ to denote the
subcube consisting of all possible assignments to variables in $L(\rho)$.
We sometimes refer to ``a random restriction $\brho \leftarrow {\cal R}_{k,n}$'' to indicate that $\brho$ is selected uniformly at random from ${\cal R}_{k,n}.$

A random restriction $\brho \samp \Rkn$ can be chosen by first picking a set
$\bK \subset [n]$ of $k$ co-ordinates to set to $\star$ and then picking $\brho_{\bar{\bK}} \in
\zo^{[n]\setminus \bK}$ uniformly at random. Often we will 
pick both $\x \in \zo^n$ and $\bK \subset [n]$ of size $k$ 
independently and uniformly at random. This is equivalent to sampling
a random restriction $\brho$ and a random point $\by$ within the subcube $C(\brho)$.

The following two theorems show that $\Iffk[f]$ captures the degree of $f_\brho$, whereas
$\sffk(f)$ captures its sensitivity.  


\begin{Thm}
\label{thm:sk}
Let $f: \zo^n \rgta \pmo$, $\brho \leftarrow \Rkn$, and $1 \leq j \leq k$. Then
\begin{equation} \label{eq:sens-moment}
\frac{\sff{j}(f)}{n^j}\approx
\frac{\sff{j}(f)}{\ff{n}{j}} \leq \Pr_{\brho \samp \Rkn} [\smax(f_\brho) \geq j] \leq  \frac{2^k\sff{j}(f) {k \choose j}}{\ff{n}{j}}
\approx 
\frac{2^k\sff{j}(f) {k \choose j}}{n^j}.
\end{equation}
\end{Thm}
\begin{proof}
Consider the bipartite graph in which the vertices $X$ on the left are all $j$-edge stars $S$ in $G_f$, the vertices $Y$ on 
the right are all restrictions $\rho \in \Rkn$, and an edge connects $S$ and $\rho$ if the star $S$ lies in the subcube $C(\rho)$ specified by the restriction $\rho$.  The desired probability
$\Pr_{\rho \in \Rkn}[ s(f_\rho) \geq j]$ is the fraction of nodes in $Y$ that are incident to at least one edge.

The number of nodes on the left is equal to 
\[
|X|=\sum_{x \in \{0,1\}^n} {s(f,x) \choose j}= {\frac {2^n  \sff{j}(f)} {j!}}.
\]
The degree of each node $S$ on the left is exactly ${n-j \choose k-j}$, since if $S$ is adjacent to $\rho$ then $j$ of the $k$ elements of $L(\rho)$ must correspond to the $j$ edge coordinates of $S$ and the other $k-j$ elements of $L(\rho)$ can be any of the $n-j$ remaining coordinates (note that the non-$\star$ coordinates of $\rho$ are completely determined by $S$).  
On the right, a restriction $\rho \in \Rkn$ is specified by a set $L(\rho)$ of $k$ live co-ordinates where $\rho_i =\star$, and
a value $\rho_i \in \zo$ for the other coordinates, so $|Y|=|\Rkn| = {n \choose k}2^{n-k}.$
We thus have
\[
\Pr_{\brho \samp \Rkn} [\smax(f_\brho) \geq j] \leq {\frac {\text{total \# of edges into $Y$}}{|Y|}} = {\frac {\left({\frac {2^n \sff{j}(f)} {j!}}  \right) \cdot {n-j \choose k-j}} { {n \choose k}2^{n-k}}} = \frac{2^k\sff{j}(f) {k \choose j}}{\ff{n}{j}}.
\]
For the lower bound, in order for $S$ to lie in $C(\rho)$ the root of $S$ must belong to $C(\rho)$ ($2^k$ choices) and all edges of $S$ must correspond to elements of $L(\rho)$ (${k \choose j}$ choices), so the maximum degree of any $\rho \in Y$
is $2^k {k \choose j}$.  Hence we have
\[
\Pr_{\brho \samp \Rkn} [\smax(f_\brho) \geq j] \geq {\frac {{\frac {\text{(total \# of edges into $Y$)}}{\text{(max degree of any $\rho \in Y$)}}}}{|Y|}} = {\frac {\left({\frac {2^n  \sff{(j)}(f)} {j!}}  \right) \cdot {n-j \choose k-j}} { 2^k {k \choose j} \cdot {n \choose k}2^{n-k}}} = \frac{\sff{j}(f)}{\ff{n}{j}},
\]
\end{proof}

\begin{Thm}
\label{thm:ik}
\footnote{The upper bound in the following theorem is essentially equivalent to Theorem 3.2 of \cite{Tal14quantum},  while the lower bound  is analogous to \cite{LMN:93}. The only difference is in the family of restrictions.}
Let $f: \zo^n \rgta \pmo$ and $\rho \samp \Rkn$. Then
\begin{equation} \label{eq:inf-moment}
\frac{\Iffk(f)}{n^k} \approx
\frac{\Iffk(f)}{\ff{n}{k}} \leq \Pr_{\brho \samp \Rkn} [\deg(f_\brho)
  = k] \leq  \frac{2^{2k-2}\Iffk(f)}{\ff{n}{k}}
  \approx \frac{2^{2k-2}\Iffk(f)}{n^k}.
\end{equation}
\end{Thm}
\begin{proof}We first fix $K \subseteq [n]$ and consider the restricted function
$f_\brho$ that results from a random choice of $\y = \brho_{\bar{K}}
\in \zo^{[n]\setminus K}$. The degree $k$ Fourier coefficient of
$f_\brho$ equals $\hat{f_\brho}(K)$ and is given by
\begin{align*}
\hat{f_\brho}(K) = \sum_{S \subset [n]\setminus K}\hat{f}(S \cup
K)\chi_S(\y).
\end{align*}
Hence we have
\begin{align*}
\Ex_{\y}[\hat{f_\brho}(K)^2] = \sum_{S \subset [n]\setminus
  K}\hat{f}(S \cup K)^2,
\end{align*}
and hence over a random choice of $\mb{K}$, we have
\begin{align}
\label{eq:eq:1}
\Ex_\brho[\hat{f_\brho}(\mb{K})^2] = \sum_{S \subset [n]}\Ex_\brho[\ind{\mb{K} \subseteq
    S}] \hat{f}(S)^2 = \sum_{S \subset [n]}\frac{\ff{|S|}{k}}{\ff{n}{k}} \hat{f}(S)^2 = \frac{\Iffk[f]}{\ff{n}{k}}.
\end{align}
Note that $\deg(f_\brho) = k$ iff $\widehat{f_\brho}(\mb{K})^2 \neq
0$. Further, when it is non-zero $\widehat{f_\brho}(\mb{K})^2$ lies in the
range $[2^{-(2k -2)},1]$, since a non-zero Fourier coefficient in a
$k$-variable Boolean function has magnitude at least $2^{-k+1}$. Hence we
have
\begin{align}
\label{eq:ineq:1}
2^{-2k +2} \Pr_\brho[\widehat{f_\brho}(\mb{K})^2 \neq 0] \leq
\Ex_\brho[\widehat{f_\brho}(\mb{K})^2]  \leq \Pr_\brho[\widehat{f_\brho}(\mb{K})^2 \neq 0]
\end{align}
which gives the desired bound when plugged into Equation (\ref{eq:eq:1}).
\end{proof}

\noindent{\bf Conjecture \ref{conj:moments} revisited: } 
\newred{An easy adaptation of the Theorem \ref{thm:ik} argument gives bounds on $\Pr_{\brho \samp \Rkn} [\deg(f_\brho) \geq j]$.}  Given these \newred{bounds}, Conjecture \ref{conj:moments} implies that for any $j \leq k$, 
\[ \Pr_{\brho \samp \Rkn} [\deg(f_\brho) \geq j] \leq a_k\Pr_{\brho \samp \Rkn} [s(f_\brho) \geq j] + o_n(1).\]
Indeed, by specifying the $o_n(1)$ \newred{term}, we can get a reformulation of Conjecture \ref{conj:moments}. 
This formulation has an intuitive interpretation: {\em gap examples exhibiting low sensitivity but high degree  are not robust to random restrictions}.  Currently, we do not know how to upper bound $\deg(f)$ by a polynomial in $s(f)$, indeed we do know of functions $f$ where $\deg(f) \geq s(f)^2$.  But the Conjecture implies that if we hit any function $f$ with a random restriction, the probability that the restriction has large degree can be bounded by the probability that it has large sensitivity. Thus the conjecture predicts that these gaps do not survive random restrictions in a rather strong sense.

\medskip 

\noindent {\bf Implications for $\acz$:} For functions with small $\acz$ circuits, a sequence of celebrated results culminating in the work of H\aa stad \cite{Hastad:86} gives upper bounds on $\Pr[\dt(f_\brho) \geq j]$. Since $\Pr[\dt(f_\brho) \geq j] \geq \Pr[\deg(f_\brho) \geq j]$, we can plug these bounds into Theorem \ref{thm:ik} to get upper bounds on the Fourier moments, and derive a statement analogous to \cite[Lemma 7]{LMN:93}, \cite[Theorem 1.1]{Tal:14tightbounds} on the Fourier concentration of functions in $\acz$. 

Similarly $\Pr[\dt(f_\brho) \geq j] \geq \Pr[s(f_\brho) \geq j]$, so via this approach Theorem \ref{thm:sk} gives upper bounds on the sensitivity moments, and hence sensitivity tail bounds for functions computed by small $\acz$ circuits. This can be viewed as an extension of \cite[Lemma 12]{LMN:93}, which bounds the average sensitivity (first moment) of such functions. For depth $2$ circuits, such tail bounds are given by the satisfiability coding lemma \cite{PPZ97}, but we believe these are the first such bounds for depth $3$ and higher. As this is not the focus of our current work, we leave the details to the interested reader.




\section{Tree sensitivity}
\label{sec:t}

In this section we study the occurrence of trees of various types in the sensitivity graph $G_f$, by defining a complexity measure called tree sensitivity. We study its relation to other complexity measures like decision tree depth.

\begin{Def}
A set $S \subseteq \zo^n$ \emph{induces a sensitive tree} $T$ in $G_f$ if 
(i) the points in $S$ induce the (non-trivial) tree $T$ in the Boolean hypercube;
(ii) every edge induced by $S$ is a sensitive edge for $f$, i.e. belongs to $E(G_f)$; and 
(iii) each induced edge belongs to a distinct co-ordinate direction.
\end{Def}

Given a fixed function $f$, a sensitive tree $T$ is completely specified
by the set $V(T)$ of its vertices. We can think of each edge $e \in
E(T)$ as being labelled by the coordinate $\ell(e) \in [n]$ along which $f$ is sensitive, so 
every edge has a distinct label. Let $\ell(T)$ denote the set of
all edge labels that occur in $T$. We refer to $|\ell(T)|$ as
the \emph{size} of $T$, and observe that it lies in $\{1,\ldots, n\}$.
We note that $|V(T)| = |\ell(T)| +1$ by the tree property. Further, any two
vertices in $V(T)$ differ on a subset of coordinates in
$\ell(T)$. Hence the set $V(T)$ lies in a subcube spanned by coordinates in
$\ell(T)$, and all points in $V(T)$ agree on all the coordinates in $\bart \eqdef [n]\setminus \ell(T)$.

\begin{Def}
\label{def:s-tree}
For $x \in \zo^n$,  the \emph{tree-sensitivity of $f$ at $x$},
denoted $\ts(f,x)$, is the maximum of $|\ell(T)|$ over all
sensitive trees $T$ such that $x \in V(T)$. We define the
tree-sensitivity of $f$ as
$\ts(f) = \max_{x \in \zo^n} \ts(f,x)
$.
\end{Def}

Note that a vertex and all its sensitive neighbors induce a
sensitive tree (which is a star). Thus one can view tree-sensitivity
as a generalization of sensitivity, and hence we have that 
$\ts(f) \geq \s(f).$
Lemma \ref{lem:gap} will show that $\ts(f)$ can in fact be exponentially
larger than both $\s(f)$ and $\dt(f)$ (the decision tree depth of $f$), and thus it cannot be upper bounded by some 
polynomial in standard measures like decision tree depth, degree, or block sensitivity. However, Theorem \ref{thm:stree-dtree}, which we  prove in the next subsection, gives a polynomial lower bound. 

\subsection{Tree sensitivity and decision tree depth}
\label{sec:ts-dt}

\red{A sensitive tree $T$ is \emph{maximal} if there does not exist any sensitive tree
$T'$ with $V(T) \subsetneq V(T')$.  In this subsection we study maximal sensitive trees using a ``shifting'' technique, introduce the notion of an ``orchard'' (a highly symmetric configuration of isomorphic sensitive trees that have been shifted in all possible ways along their insensitive coordinates), and use these notions to prove Theorem \ref{thm:stree-dtree}, which lower bounds tree sensitivity by square root of decision tree depth.}

The \emph{support} of a vector $v \in \zo^n$, denoted $\supp(v)$, is the set $\{i \in [n]: v_i=1\}.$
For $x,v \in \zo^n$, $x \oplus v$ denotes the coordinatewise xor. Given a set $S \subseteq \zo^n$, let $S \oplus v = \{x \oplus
v: x \in S\}$.

\begin{Def}
\label{def:shift}
Let $v$ be a vector supported on $\bart$ where $T$ is a sensitive tree in $G_f$. We say that $T$ can be
\emph{shifted} by $v$ if $f(x) = f(x \oplus v)$ for all $x \in V(T)$. 
\end{Def}

If $T$ can be shifted by $v$ then $V(T) \oplus v$ also induces a sensitive tree
which we denote by $T \oplus v$.  Mapping $x$ to $x \oplus v$  gives
an isomorphism between $T$ and $T \oplus v$ which preserves both
adjacency and edge labels, and in particular we have $\ell(T \oplus v) = \ell(T)$.

We have the following  characterization of maximality (both directions follow easily
from the definitions of maximality and of shifting by the unit basis vector $e_i$):
\begin{Lem}
\label{lem:maximal}
A sensitive tree $T$ is maximal if and only if it can be shifted by 
$e_i$ for all $i \in \bart$ (equivalently, if none of the vertices in $V(T)$ is sensitive to
any coordinate in $\bart$).
\end{Lem}

The notion of maximality allows for a ``win-win'' analysis of
sensitive trees: for each co-ordinate $i \in \bart$, we can either
increase the size of the tree by adding an edge in direction $i$, or
we can shift by $e_i$ to get an isomorphic copy of the tree. Repeating
this naturally leads to the following definition.

\begin{Def}
\label{def:orchard}
Let $T$ be a sensitive tree that can be shifted by every $v$
supported on $\bart$. We refer to the set of all such trees $F =
\{T \oplus v\}$ as an \emph{orchard}, and we say that $T $ belongs
to the orchard $F$. 
\end{Def}

An orchard guarantees the existence of $2^{n - \ell(T)}$ trees that are
isomorphic to $T$ in $G_f$. It is {\em a priori} unclear that orchards exist in $G_f$.
The following simple but key lemma proves their existence.

\begin{Lem}
\label{lem:main-tree}
Let $T$ be a sensitive tree. Either $T$ belongs to an orchard, or there
exists a shift $T \oplus v$ of $T$ which is not maximal.
\end{Lem}
\begin{Proof}
Assume the tree $T$ does not belong to an orchard. 
Pick the smallest weight vector $v'$ supported on $\bart$ such
that $T$ cannot be shifted by $v'$ (if there is more than one such vector any one will do). 
Since $T$ can trivially be
shifted by  $0^n$, we have $\wt(v') \geq 1$. Pick any
co-ordinate $i \in \supp(v')$, and let $v = v' \oplus e_i$ so that
$\wt(v) = \wt(v') -1$. By our choice of $v'$, $T$ can be shifted 
by $v$, but not by $v' = v \oplus e_i$. This implies that
there exists $x \in V(T)$ so that
$f(x)  =f(x \oplus v) \neq f(x \oplus v')$,
hence $T \oplus v$ is not maximal.
\end{Proof}

This lemma directly implies the existence of orchards for every $G_f$:

\begin{Cor}
\label{cor:2}
Every sensitive tree $T$ where $|\red{\ell(T)}| =\ts(f)$ belongs to an orchard.
\end{Cor}

The lemma also gives the following intersection property for orchards. Since
any two trees in an orchard $F$ are isomorphic, we can define $\ell(F) = \ell(T)$ to
be the set of edge labels for any tree $T \in F$. 

\begin{Lem}
\label{lem:int-orchard}
Let $F_1$ and $F_2$ be orchards. Then $\ell(F_1) \cap \ell(F_2) \neq \emptyset$.
\end{Lem}
\begin{Proof}
Assume for contradiction that $\ell(F_1)$ and $\ell(F_2)$ are
disjoint. We choose trees $T_1 \in F_1$ and $T_2 \in F_2$, and $x \in
V(T_1), y \in V(T_2)$ such that $f(x) =1$ and $f(y) =-1$. Now define $z \in \zo^n$ where
$z_i$ equals $ x_i$ if $ i \in \ell(T_1)$ and $z_i$ equals $y_i$ otherwise.
Since $z$ agrees with $x$ on $\ell(T_1) = \ell(F_1)$, it can be obtained by
shifting $x$ by $z \oplus x$ which is supported on
$\overline{\ell(T_1)}$. Since $T_1$ belongs to an orchard, we get
$f(z) = f(x) = 1$.  However, we also have that  $z_i = y_i$ for all $i \in \ell(T_2)$. Hence by similar
reasoning, $f(z) = f(y) = -1$, which is a contradiction.
\end{Proof}

We use this intersection property to lower bound tree sensitivity in terms of decision tree depth, via an argument similar to other upper bounds on $\dt(f)$ (such as the well known \cite{BlumImpagliazzo:87,Tardos89,HH91} quadratic upper bound on $\dt(f)$ in terms of certificate complexity).  

\begin{Thm}
\label{thm:stree-dtree}
For any Boolean function $f: \zo^n \rgta \pmo$, we have 
$\ts(f) \geq \sqrt{2\dt(f)} -1.$
\end{Thm}
\begin{Proof}
We construct a decision tree for $f$ by iterating the following step
until we are left with a constant function at each leaf: at the current node in the decision tree, pick the largest sensitive
tree $T$ in the (restricted) function and read all the variables in $\ell(T)$.

Let $k$ be the largest number of iterations before we terminate, taken
over all paths in the decision tree.  Fix a path that achieves
$k$ iterations and
let $f_i$ be the restriction of $f$ that is obtained, at the end of the $i$-th
iteration (and let $f_0 =f$).
We claim that $\ts(f_i) \leq \ts(f) - i$. Note that if $f_i$
is not constant then $\ts(f_i) \geq 1$, hence this claim implies
that $k \leq \ts(f)$.

It suffices to prove the case $i =1$, since we can then apply the same
argument repeatedly. Consider all trees in $f_0 = f$ of
size $\ts(f)$. Each of them occurs in an orchard by Corollary
\ref{cor:2} and by Lemma \ref{lem:int-orchard} any two of them share
at least one
variable. Hence when we read all the variables in some tree $T$,
we restrict at least one variable in every tree of size
$\ts(f)$, reducing the size by at least $1$. The size of the other trees
cannot increase after restriction, since $G_{f_1}$ is an induced
subgraph of $G_{f}$. Hence all the sensitive trees in $f_1$ have
size at most $\ts(f) -1$.

It follows that overall we can bound the depth of the resulting decision tree by
\begin{align*}
\dt(f) \leq \sum_{i=1}^k\ts(f_{i-1}) \leq \sum_{i=1}^k(\ts(f) - (i-1))
\leq \frac{\ts(f)(\ts(f) +1)}{2}.
\end{align*}
\end{Proof}

It is natural to ask whether $\ts(f)$ is polynomially related to $\dt(f)$
and other standard complexity measures. Lemma \ref{lem:gap} in Section \ref{sec:ex} gives an example of a function on $n$ variables where $\dt(f) = \log(n+1)$ whereas $\ts(f) = n$.  In the other direction, it is likely that the bound in Theorem
\ref{thm:stree-dtree} can be improved further. We conjecture that the
following bound should hold:

\begin{Conj}
\label{conj:1}
For any Boolean function $f: \zo^n \rgta \pmo$, we have
$\ts(f) \geq \dt(f)$ (and hence $\ts(f) \geq \deg(f)$).
\end{Conj}
In addition to being a natural question by itself, we will show in Section \ref{sec:ts-app} that Conjecture
\ref{conj:1} would have interesting consequences via the switching
lemma in Section \ref{sec:ts}.




\subsection{Tree Sensitivity under Random Restrictions} 
\label{sec:ts}

In this subsection we show that the probability of a random restriction of $f$ having large
tree sensitivity is both upper and lower bounded by suitable
sensitivity moments of $f$. 

\begin{Thm}
\label{thm:ts}
Let $f: \zo^n \rgta \pmo$, $\brho \leftarrow \Rkn$ and $1 \leq j \leq k$. Then we have
\begin{align*} 
\frac{\sff{j}(f)}{n^j} 
\approx
\frac{\sff{j}(f)}{\ff{n}{j}} 
\leq 
\Pr_{\brho \leftarrow \Rkn}[ \ts(f_\brho) \geq j]  
\leq 
\frac{{k\choose j} 2^{k +2j} s^{j}(f)}{{n \choose j}}
\approx
2^k\frac{(4k)^{j} s^{j}(f)}{n^j}.
\end{align*}
\end{Thm}

The lower bound follows from the fact that $\ts(f) \geq s(f)$ and Theorem \ref{thm:sk}. The key ingredient in the upper bound is  Sidorenko's theorem \cite{Sidorenko:94}, which bounds the number of homomorphisms from a tree $T$ with $j$ edges to a graph $G$ to in terms of the $j^{th}$ degree moment of $G$. For a formal statement of Sidorenko's theorems, we refer the reader to \cite{Sidorenko:94,CL14}. Below, we state the result we will use in our language. 
We also present an elegant proof due to Levin and Peres \cite{LevinPeres16} which seems considerably simpler than the known proofs of Sidorenko's theorem (though the lemma follows directly from that theorem). 

\begin{Lem} \cite{LevinPeres16} \label{lem:peres}
Let $\calS_j$ denote the set of sensitive trees of size $j$ in $G_f$. Then we have that
\[ |\calS_j| \leq 4^j \sum_{x \in \zo^n}s(f,x)^j. \]
\end{Lem}
\begin{proof}
We consider the set $\calT$ of all rooted unlabelled trees with $j$ edges. It is known that $|\calT| \leq 4^j$, indeed it equals the $j^{th}$ Catalan number. Each tree $t \in \calT$ has $j +1$ vertices. We label them $\{0,\ldots,j\}$ where $0$ is the root, and we label the remaning vertices using a breadth first search. This lets us define $p_t:\{1,\ldots,j\} \to \{0,\ldots,j-1\}$ where $p_t(i) < i$ is the parent of vertex $i$.  Let $\calS(t)$ denote the set of sensitive trees $T \in G_f$ whose adjacency structure is given by $t$.


For conciseness let us write $\stot(f)$ to denote $\sum_{x \in \{0,1\}^n} s(f,x).$
Let $\calD$ denote the distribution on $\zo^n$ where for each $x \in \{0,1\}^n$,
\[\Pr_{\bx \leftarrow \calD}[\bx=x] = \frac{s(f,x)}{\stot(f)}.\] 
Note that $\calD$ is supported only on vertices where $s(f,x) \geq 1$. Further $\calD$ is a stationary distribution for the simple random walk on $G_f$: if we sample a vertex from $\calD$ and then walk to a random neighbor, it is also distributed according to $\calD$. 

Fix a tree $t \in \calT$ and consider a random walk on $G_f$ which is the following vector $\bX =(\bX_0,\ldots,\bX_j)$ of random variables:
\begin{itemize}
\item We sample $\bX_0$ from $\zo^n$ according to $\calD$. 
\item For $i \geq 1$,  let $\bX_i$ be a a random neighbor of $\bX_{i'}$ in $G_f$ where $i' = p_t(i) <i$.
\end{itemize}
Note that every $\bX_i$ is distributed according to $\calD$. The vector $\bX = (\bX_0,\ldots,\bX_j)$ is such that $(\bX_{i},\bX_{p_t(i)}) \in E(G_f)$, but it might contain repeated vertices and edge labels (indeed, this proof bounds the number of homomorphisms from $G_f$ to $t$).  

A vector $x = (x_0,\ldots,x_j) \in \bkts{\zo^n}^{j+1}$ will be sampled with probability
\begin{align*} 
\Pr[\bX = x] & = \Pr[\bX_0 =x_0]\prod_{i=1}^j\Pr[\bX_i = x_i|\bX_0,\ldots,\bX_{i-1}]\\
& = \frac{s(f,x_0)}{\sum_{x \in \zo^n}s(f,x)}\prod_{i=0}^{j-1}\frac{1}{s(f,x_i)}\notag \\
 & = \frac{1}{\sum_{x \in \zo^n} s(f,x)}\prod_{i=1}^{j-1}\frac{1}{s(f,x_i)}. \end{align*}
 
Clearly $\calS(t)$ lies in the support of $\bX$, hence
\begin{align}
|\calS(t)| & \leq \supp(\bX)\notag\\ 
& \leq \E_\bX\sbkts{\frac{1}{\Pr[\bX= x]}}\notag\\ 
& \leq \E_{\bX}\sbkts{\sum_{x \in \zo^n}s(f,x) \prod_{i=1}^{j-1}s(f,\bX_i)}\notag\\
& = \stot(f)  \E_{\bX}\sbkts{\prod_{i=1}^{j-1}s(f,\bX_i)}\notag\\
& \leq \stot(f)  \E_{\bX}\sbkts{\frac{\sum_{i=1}^{j-1}{s(f,\bX_i)^{j-1}}}{j-1}}\tag{AM-GM Inequality}\\
& = \stot(f)  \E_{\bY \sim \calD}\sbkts{s(f,\bY)^{j-1}}\label{eq:sidorenko}
\end{align}
where the last equality holds by linearity of expectation and the fact that all the $\bX_i$'s are identically distributed. 
We bound the moment under $\calD$ as follows:
\begin{align*}
\E_{\bY \sim \calD}\sbkts{s(f,\bY)^{j-1}} & \leq \sum_{y \in \zo^n} \Pr[\bY =y]s(f,y)^{j-1}\\
&= \sum_{y \in \zo^n} \frac{s(f,y)}{\stot(f)}s(f,y)^{j-1}\\
&= \frac{\sum_{y \in \zo^n} s(f,y)^j}{\stot(f)}.
\end{align*}

Plugging this back into Equation \eqref{eq:sidorenko} gives 
\begin{align*}
|\calS(t)| \leq \sum_{y \in \zo^n}s(f,y)^j
\end{align*}
Summing over all possibilities for $t$, we get
\[ |\calS_j| \leq \sum_{t \in \calT}|\calS(t)| \leq 4^j\sum_{y \in \zo^n}s(f,y)^j.\]
\end{proof}

Theorem \ref{thm:ts} now follows from an argument similar to Theorem \ref{thm:sk}. 
\begin{proof}[Proof of Theorem \ref{thm:ts}]
The lower bound follows from (the lower bound in) Theorem \ref{thm:sk} and the observation that $\ts(f_\rho) \geq
s(f_\rho)$. We now prove the upper bound.

Similar to Theorem \ref{thm:sk}, consider the bipartite graph where the LHS is the set $\calS_j$ of all sensitive
trees $T$ of size $j$ in $G_f$,  the RHS is the set $\Rkn$ of all restrictions $\rho$, and $(T,\rho)$ is an edge if the tree
$T$ lies in the subcube $C(\rho)$ specified by the restriction $\rho$.  The desired probability
$\Pr_{\brho \in \Rkn}[ \ts(f_\brho) \geq j]$ is the fraction of nodes in $\Rkn$ that are incident to at least one edge.

We first bound the degree of each vertex on the left.  To have $T$ lying in $C(\rho)$, 
\begin{itemize}
\item The edge labels of $T$ must be live variables for $\rho$. 
\item The values $\rho_i$ for the fixed coordinates $i \in [n]\setminus L(\rho)$ must be consistent with the values in $V(T)$. 
\end{itemize}
The only choice is of the $(k -j)$ remaining live coordinates. Hence $T \in C(\rho)$ for at most ${n -j \choose k -j}$ values of $\rho$ corresponding to choices of the remaining live variables. 

The number of vertices in $\calS_j$ is bounded using Lemma \ref{lem:peres} by 
\[
|\calS_j| \leq 4^j \sum_{x \in \zo^n}s(f,x)^j = 4^j2^ns^j(f),
\]
so the total number of edges is at most 
\[
{n -j \choose k - j}2^n 4^js^{j}(f).
\]
A restriction $\rho \in \Rkn$ is specified by a set $L(\rho)$ of $k$ live co-ordinates where $\rho_i =\star$, and
a value $\rho_i \in \zo$ for the other coordinates, and hence
\[
|\Rkn| = {n \choose k}2^{n-k}.
\]
Recall that $\ts(f_\rho) \geq j$ iff $C(\rho)$ contains some tree from $\calS_j$.
Hence the fraction of restrictions $\rho$ that have an edge incident to them is
\begin{align*}
\Pr_{\rho \in \Rkn}[ \ts(f_\rho) \geq j] & \leq  \frac{{n -j
    \choose k - j}2^n4^js^{j}(f)}{{n \choose k}2^{n-k}}
\leq \frac{{k\choose j} 2^{k +2j} s^{j}(f)}{{n \choose j}}.
\end{align*}
\end{proof}

\subsection{An approach to Conjecture \ref{conj:moments}}
\label{sec:ts-app}

By combining Theorems \ref{thm:sk}, \ref{thm:stree-dtree} and \ref{thm:ts}, we get
upper and lower bounds on the probability that a random restriction of
a function has large decision tree depth in terms of its sensitivity
moments.

\begin{Cor}
\label{cor:sf}
Let $f: \zo^n \rgta \pmo$, $\brho \sim \Rkn$ and $1 \leq j \leq k$. Then
\begin{align*} 
\frac{\sff{j}(f)}{n^j} \approx 
\frac{\sff{j}(f)}{\ff{n}{j}} \leq  \Pr_{\brho \leftarrow \Rkn}[ \dt(f_\brho) \geq j]  
\leq 8^k\frac{{k\choose \sqrt{2j}-1} s^{\sqrt{2j} -1}(f)}{n^{\sqrt{2j} - 1}}
\approx \frac{8^k k^{\sqrt{2j} -1}s^{\sqrt{2j} -1}(f))}{n^{\sqrt{2j} - 1}}.
\end{align*}
\end{Cor}

Note that the denominator in the lower bound is $n^{\Omega(j)}$ but for
the upper bound, it is $n^{\Omega(\sqrt{j})}$. This quadratic gap
comes from Theorem \ref{thm:stree-dtree}. However, if Conjecture
\ref{conj:1} stating that $\ts(f) \geq \dt(f)$ were true, it would
imply the following sharper upper bound.

\begin{Cor}
\label{cor:sf+conj1}
Let $f: \zo^n \rgta \pmo$, $\rho \sim \Rkn$ and $1 \leq j \leq k$. If
Conjecture \ref{conj:1} holds (in the stronger form that $\ts(f) \geq \dt(f)$), then
\begin{align*} 
\Pr_{\brho \leftarrow \Rkn}[ \dt(f_\brho) \geq j]   \leq \frac{{k\choose j} 2^{k +2j} s^{j}(f)}{{n \choose j}},
\end{align*}
and if Conjecture \ref{conj:1} holds in the weaker form that $\ts(f) \geq \deg(f)$, then
\begin{align*} 
\Pr_{\brho \leftarrow \Rkn}[ \deg(f_\brho) \geq j]   \leq \frac{{k\choose j} 2^{k +2j} s^{j}(f)}{{n \choose j}}.
\end{align*}
\end{Cor}

The dependence on $n$ here matches that in  the lower bound of
Corollary \ref{cor:sf}.  Conjecture \ref{conj:moments} follows from this as an easy consequence. 
\ignore{(indeed showing $\ts(f) \geq \deg(f)$ rather than Conjecture \ref{conj:1} suffices).}

\begin{Cor}
\label{cor:sf+conj2}
Conjecture \ref{conj:1} (in fact, the weaker form that $\ts(f) \geq \deg(f)$) implies Conjecture \ref{conj:moments}.
\end{Cor}
\begin{Proof}
We will prove that
$\Iffk(f) \leq 8^k k! s^{k}(f).$
Let $\brho \leftarrow \Rkn$ and consider the event that $\deg(f_\brho) =
k$. By Theorem \ref{thm:ik}, we can lower bound this probability in
terms of the Fourier moments of $f$ as
\[
\frac{\Iffk(f)}{\ff{n}{k}} \leq \Pr_{\brho \samp \Rkn} [\deg(f_\brho)= k].
\]
To upper bound it, by Corollary \ref{cor:sf+conj1}, if the weaker form of Conjecture \ref{conj:1} holds, then we have
\[\Pr_{\brho \leftarrow \Rkn}[ \deg(f_\brho) \geq k] \ignore{ \leq  \Pr_{\brho \leftarrow \Rkn}[ \dt(f_\brho) \geq k]   }
\leq \frac{2^{3k} s^{k}(f)}{{n \choose k}}. \]

The claim follows by comparing the upper and lower bounds.
\end{Proof}

For $k =3, 4$, it is an easy exercise to verify that $\deg(f_\rho) = k$ implies $\ts(f_\rho) \geq k$.\ignore{\anote{I could verify the case $k=3,4$ with degree instead of decision-tree complexity (which I couldn't), since for degree I can assume without loss of generality that $n=k$, and for decision tree I could not necessarily. For $k=5$, I ran an exhaustive search and verified that $\deg(f)=5$ implies $\ts_2(f)\ge 5$. In general we can consider stating Conjecture~\ref{conj:1} in terms of $\ts(f) \ge \deg(f)$ as it is enough to imply Conjecture~\ref{conj:moments}.}} This implies that Conjecture \ref{conj:moments} holds for $k =3, 4$.

\ignore{\pnote{Changed title}}
\section{Fourier concentration for low-sensitivity functions}
\label{sec:pw}

\subsection{Proper Walks}
\label{sec:def-pw}

Since $s^j(f) \leq (s(f))^j$ for all $j$, one can trivially bound the sensitivity moments of a function in terms of its max sensitivity.  Hence 
Corollaries \ref{cor:sf+conj1} and \ref{cor:sf+conj2} show that under Conjecture \ref{conj:1}, low sensitivity functions simplify under random restrictions. In this section we prove this unconditionally. The key ingredient is a relaxation of sensitive trees that we call \emph{proper walks}.

A \emph{walk} $W$ in the $n$-dimensional Boolean cube is a sequence of vertices
$(w_0,w_1,\ldots,w_t)$ such that $w_i$ and $w_{i+1}$ are at
Hamming distance precisely $1$.  We allow walk to backtrack and visit vertices more than once.  
We say that $t$ is the \emph{length} of such a walk.

Let $\ell(W) \subseteq [n]$ denote the set of coordinates that are flipped by  walk $W$. We define $k = |\ell(W)|$ to be the \emph{dimension} of the walk.
We order the coordinates in $\ell(W)$ as $\ell_1,\ldots,\ell_k$ according to the order in which they are first flipped.  
For each $\ell_i \in \ell(W)$, let $x_i$ denote the first vertex in $W$ at which we flip coordinate $i$.

\begin{Def}
A walk $W$ is a \emph{proper walk} for a Boolean function $f: \zo^n \rgta \pmo$ if for each $\ell_i \in \ell(W)$, 
the vertex $x_i$ is sensitive to $\ell_i$.
\end{Def}

Thus a walk is proper for $f$ if the first edge flipped along a new coordinate direction is always sensitive.
This implies that while walking from $x_i$ to $x_{i+1}$, we are only allowed to flip a subset of the coordinates $\{\ell_1,\ldots,\ell_i\}$, hence $\supp(x_i \oplus x_{i+1}) \subseteq \{\ell_1,\ldots, \ell_i\}$. Hence if there is a proper walk of dimension $k$ then there is one of length at most $k(k+1)/2$, by choosing a shortest path between $x_i$ and $x_{i+1}$ for each $i$.

In studying proper walks, it is natural to try to maximize the dimension and minimize the length. We first focus on the former. The following lemma states that the obvious necessary condition for the existence of an $n$-dimensional walk is in fact also sufficient:

\begin{Lem}
Every Boolean function $f:\zo^n \rgta \pmo$ that depends on all $n$
coordinates has a proper walk of dimension $n$.
\end{Lem}
\begin{Proof}
Pick $\ell_1 \in [n]$ arbitrarily and let $x_1$ be any vertex in $\zo^n$ which is sensitive to
coordinate $\ell_1$.  Let $1  \leq i \leq n$. Inductively we assume we have picked coordinates $L =\{\ell_1,\ldots,\ell_i\}$ and points $X = \{x_1,\ldots,x_i\}$ so that for every $j \leq i$, 
\begin{enumerate}
\item $x_j$ is sensitive to $\ell_j$.
\item For $j \geq 2$, $\supp(x_{j-1} \oplus x_{j}) \subseteq \{\ell_1,\ldots,\ell_{j-1}\}$. 
\end{enumerate}
If we visit $x_1,\dots,x_i$ in that order and walk from each $x_j$ to $x_{j+1}$ along a shortest path, the resulting walk is a proper walk for $f$.
Let $C$ be the subcube that spans the dimensions in $L$ and contains $X$.

{\bf Case 1:} Some vertex in $C$ is sensitive to a coordinate outside
of $L$. Name this vertex $x_{i+1}$ and the sensitive
co-ordinate $\ell_{i+1}$, and add them to $X$ and $L$ repectively. Note that $x_{i} \oplus x_{i+1}$ is indeed
supported on $\{\ell_1,\ldots,\ell_i\}$, so both conditions (1) and (2) are met. 

{\bf Case 2:} No vertex in $C$ is sensitive to a coordinate outside
$L$. So for any co-ordinate $j \not\in L$, we have
$f(x) = f(x \oplus e_j)$.
But this means that the set of points $X \oplus e_j$ and co-ordinates $L$ also satify the inductive hypothesis (specifically conditions (1) and (2) above).

Let $d$ denote the Hamming distance from $C$ to the closest
vertex which is sensitive to some coordinate outside $L$.
Let $z$ denote one such closest vertex to $C$ (there could be many)
and pick any coordinate $j$ in which $z$ differs from the closest point 
in $C$.  If we replace $X$ by  $X \oplus e_j$, the Hamming
distance to $z$ has decreased to $d-1$. We can repeat this till the
Hamming distance drops to $0$, which  puts us in Case (1).
\end{Proof}

Given this result, it is natural to try to find full dimensional walks of the smallest possible length. The length of the walk constructed above is bounded by $\sum_{i=1}^n(i-1) \leq n^2/2$. Lemma \ref{lem:ex2} in Section \ref{sec:ex} gives an example showing that this is tight up to constants. So while we cannot improve the bound in general, we are interested in the case of functions with large decision tree complexity, where the following observation suggests that better bounds should be possible.

\begin{Lem}
\label{lem:ts-gives-pw}
If $\ts(f) =n$, then $f$ has a proper walk of dimension $n$ and length \red{$2n$}.
\end{Lem}

The proof is by performing a 
traversal on a sensitive
tree of dimension $n$, starting and ending at the root, going over each edge twice. Thus if Conjecture \ref{conj:1} were true,
it would imply that functions requiring full decision tree depth have
proper walks of length $O(n)$. We now give an
unconditional proof of this result (we will use it as an essential ingredient in our ``switching lemma'' later).
\begin{Thm}
\label{thm:dt-gives-pw}
If $\dt(f) = n$, then $f$ has a proper walk of dimension $n$ and length at most $3n$.
\end{Thm}
\begin{Proof}
The proof is by induction on $n$.
The base case $n = 2$ is trivial since in this case there exists a proper walk of length
$2$. Assume the claim holds for all $n' < n$.
Let $f$ be a function where $\dt(f) =n$. If $\ts(f) = n$ we are done
by Lemma \ref{lem:ts-gives-pw}, so we assume that $\ts(f) = m < n$.
By Corollary \ref{cor:2}, there is an orchard $\{T \oplus v\}$ of sensitive trees 
where $\dim(T) = m$. Assume by relabeling that $\ell(T) = \{1,\ldots,m\}$.

Since $\dt(f) = n$, there exists a setting $t_1,\ldots,t_m$ of
variables in $[m]$ such that the restriction $f' = f|_{x_1 =t_1,\ldots,x_m =t_m}$
on $n' = n - m$ variables satisfies $\dt(f') = n - m$. By the
inductive hypothesis, there exists a proper walk in $f'$ of dimension
$n-m$ and length $3(n - m)$ in the subcube $x_1 =t_1,\ldots,x_m =t_m$
which starts at some vertex $s' =
(t_1,\ldots,t_m,s'_{m+1},\ldots,s'_n)$ and ends at some vertex $t' = (t_1,\ldots,t_m,t'_{m+1},\ldots,t'_n)$,
which flips all coordinates in $[n]\setminus[m]$.

Consider the tree $T \oplus v$ in the orchard such that the coordinates of $V(T \oplus v)$ in
$[n]\setminus [m]$  agree with $s'$. Our walk can be divided into three phases:

\begin{enumerate}
\item By Lemma \ref{lem:ts-gives-pw}, we can visit every vertex in $T \oplus v$ using a proper walk of length $2m$ that only
uses edges in $[m]$. 
Assume that this walk starts and ends at $r$.
By our choice of $v$ we have that 
$(r_{m+1},\ldots,r_n) =  (s'_{m+1},\ldots,s'_n)$.

\item From $r$, we then walk to the vertex $s' = (t_1,\ldots,t_m,s'_{m+1},\ldots,s'_n)$.
This only requires flipping bits in $[m]$, so it keeps the walk proper and adds only at most $m$ to its length.

\item The inductive hypothesis applied to $f'$ allows us to construct a proper walk from $s'$ to $t'$
that only walks along edges in $[n]\setminus [m]$ and has length at most $3(n -m)$.
\end{enumerate}

Thus the total length of the walk is at most $2m + m  + 3(n - m) =  3n$.
\end{Proof}




\subsection{Random restrictions of low sensitivity functions}
\label{sec:pw-rr}
In this section we prove our ``switching lemma for low-sensitivity functions'', Lemma \ref{lem:switching encoding}.  The high-level idea is to count the number of (short) proper walks that arise when using Theorem \ref{thm:dt-gives-pw}. This allows us to upper bound the number of restrictions $\rho$ for which $f_{\rho}$ has full decision tree (since each such restriction yields a short proper walk by Theorem~\ref{thm:dt-gives-pw}).
We follow Razborov's proof strategy for the switching lemma \cite{Raz95}. In order to bound the number of walks, we encode each walk using a short description and then show that this encoding is a bijection.
Since the encoding is bijective,  the number of walks is at most the number of encodings and we get the required upper bound.

%
%
%
%

\begin{Lem}
\label{lem:switching encoding}
Let $f:\zo^n \rgta \pmo$. Then
\begin{align*}
\Pr_{\brho \leftarrow \Rkn}[ \dt(f_\brho) =k] \leq
\frac{(32 s(f))^k}{\binom{n}{k}}.
\end{align*}
\end{Lem}

\begin{Proof}
	We will prove the Lemma via an encoding scheme. Let $S = \{\rho \in \Rkn: \dt(f_\rho) = k\}$. We shall prove that $|S| \le 2^n \cdot (16  s(f))^k$. This will complete the proof since 
	\[\Pr_{\brho \leftarrow \Rkn}[ \dt(f_\brho) =k] = \frac{|S|}{|\Rkn|} = \frac{2^n \cdot (16  s(f))^k}{2^{n-k} \cdot \binom{n}{k}} = \frac{(32  s(f))^k}{\binom{n}{k}}.\]
	
	For shorthand, let $s = s(f)$. We define an encoding  $$E: S \to \{0,1\}^n \times \{0,1\}^k \times \{0,1\}^{2k} \times \{0,1\}^k \times [s]^{k}\;,$$ and show that $E$ is a bijection, hence  $|S| \le 2^n \cdot 2^{4k} \cdot s^k$. 
	Given a restriction $\rho \in S$, let $W = v_0, v_1, \ldots, v_{3k}$ be the proper walk defined in the proof of Theorem~\ref{thm:dt-gives-pw}.
	We encode $W$ to $(v_0, 1_{K}, b, c, \beta)$ where $K \subseteq [k]$, $b \in \{0,1\}^{2k}$, $c \in \{0,1\}^k$ and $\beta\in [s]^k$ are defined next.

	We open up the recursive argument from Theorem~\ref{thm:dt-gives-pw}. The walk $W$ consists of $t\le k$ phases. Each phase $i \in [t]$ consists of a traversal over a sensitive tree $T_i$ with $k_i$ edges, and then a walk to a point in the minimal cube containing $T_i$. The root of $T_{i+1}$ is the end point of the walk of  phase $i$.

	\paragraph{Encoding the tree sizes.} We encode using a binary string of length $k$ the numbers $(k_1, \ldots, k_t)$. This can be done by letting $K = \{k'_1, k'_2, \ldots, k'_t\}$ where $k'_i = k_1 + \ldots + k_i$ and taking the indicator vector of $K$. Since all $k_i$-s are nonnegative and $k_1 + \ldots + k_t = k$, we get that all $k'_1, \ldots, k'_t$ are distinct numbers between $1$ and $k$. Thus, the set $K$ is a subset of $[k]$ and may be encoded using $k$ bits.
	
\paragraph{Encoding each tree.}
In phase $i \in \{1,\ldots, t\}$, given the initial node of the phase, which is the root of $T_i$, denoted $r_i$, we show how to encode $T_i$ using $2k_i$ bits in addition to $k_i$ numbers in $[s]$ (recall that $k_i$ is the number of edges in $T_i$).

Take a traversal over the tree $T_i$ starting at the root and finishing at the root. The length of this walk is $2k_i$ since we go on each edge once in each direction.
We encode the walk as a binary string of length $2k_i$ in addition to a sequence in $[s]^{k_i}$.
For the walk $r_i = u_0, u_1, \ldots, u_{2k_i}$, define the encoding $(b^{(i)}, \beta^{(i)}) \in \{0,1\}^{2k_i} \times[s]^{k_i}$ as follows. 
Initialize $\beta^{(i)}$ to be the empty sequence, and $b^{(i)}$ to be the all zeros sequence $0^{k_i}$.
For $j=1,\ldots, 2k_i$, either $u_{j}$ is the father of $u_{j-1}$ in the sensitive tree, or $u_j$ is a child of $u_{j-1}$.
Put $b^{(i)}_j = 1$ iff  $u_{j}$ is a child of $u_{j-1}$.
In such a case, let $n_j \in [s]$ be the index of $u_{j}$ among all sensitive neighbors of $u_{j-1}$ in the hypercube, according to some canonical order on $\{0,1\}^n$, and append $n_j$ to the sequence $\beta^{(i)}$.

After finishing the walk, let $\ell_1 <\ell_2 \ldots < \ell_{k_i}$ be the coordinates that the sensitive tree $T_i$ changes (i.e., $\ell(T_i)$).
We encode using $c^{(i)} \in \{0,1\}^{k_i}$ the walk inside the minimal cube containing $T_i$ by setting $c^{(i)}_j = 1$ iff the $\ell_j$ coordinate should be flipped during the walk, for $j = 1, \ldots, k_i$.

Finally, after finishing the $t$ phases, we let $b = b^{(1)} \circ \ldots \circ b^{(t)}$,  $\beta =\beta^{(1)}  \circ \ldots \circ \beta^{(t)}$ and $c = c^{(1)} \circ \ldots \circ c^{(t)}$ be the concatenation of the substrings defined for each phase.

\paragraph{Decoding.} We show that a decoder can uniquely recover the proper walk from the encoding. This will allow us to show that $E$ is a bijection.
First, the decoder knows $k_1, \ldots, k_t$ (and $t$) since it can decode the set $K = \{k'_1, \ldots, k'_t\}$ that determines $k_1, \ldots, k_t$ by $k_1 = k'_1$ and $k_i = k'_i - k'_{i-1}$ for $i=2,\ldots, t$.

For $i=1, \ldots, t$ we show that assuming the decoder knows the starting position of the walk in phase $i$, i.e., $r_i$, it decodes phase $i$ successfully.
Since by the end of phase $i$ we reach $r_{i+1}$ this shows that the entire decoding procedure works.

Given the tree sizes $k_1,\dots,k_t$, the decoder may identify the substrings $b^{(i)}\in \{0,1\}^{2k_i}$, $\beta^{(i)}\in [s]^{k_i}$ and $c^{(i)} \in \{0,1\}^{k_i}$ inside $b$, $\beta$ and $c$ respectively.
To follow the traversal in the sensitive tree $T_i$ the decoder reads bits from $b^{(i)}$ indicating whether one should go to a child of the current node or move back to its parent. In the case that the traversal goes to a child of the current node, the next symbol from $\beta^{(i)}$ is read, indicating which one of the sensitive neighbors of the current node is this child.

After reading $2k_i$ bits from $b$ and $k_i$ symbols from $\beta$, the decoder finishes reconstructing the tree traversal of $T_i$. Thus, it identifies correctly the coordinates $\ell_1 < \ldots < \ell_{k_i}$ as the sensitive coordinates in the tree $T_i$. Next, the decoder walks along the path defined by $c^{(i)}$, flipping the $\ell_j$ coordinate iff $c^{(i)}_{j} = 1$ for $j = 1, \ldots, k_i$. This completes the decoding of phase $i$.

\paragraph{$E$ is a bijection.} Let $\rho \in S$. Given $E(\rho) = (x_0, 1_K, b, c, \beta)$ the decoder finds a proper walk $W$ with $|\ell(W)|=k$, that is contained in the subcube defined by $\rho$. Thus, the minimal subcube containing the walk $W$ uniquely determines the restriction $\rho$, and we get that $E$ is bijective.
\end{Proof}

We note that one can prove a similar bound for $\Pr_{\rho \in \Rkn}[\dt(f_\rho) \geq j]$; here we have presented only the case $j = k$ both because it is simpler and because it suffices for the concentration results in Section \ref{sec:fourier-conc}.

We would like to replace the $(s(f))^k$ term with $s^k(f)$, the
$k^{th}$ sensitivity moment. The above proof does not seem to
generalize to that case, because we do not have an analogue of Sidorenko's result on trees for proper walks.

\subsection{Fourier tails of low sensitivity functions}
\label{sec:fourier-conc}

We have the necessary pieces in place to give an upper bound on $\Iffk[f]$:

\begin{Lem} \label{lem:Iffk-upper-bound}
For every $f: \{0,1\}^n \rta \pmo$ and every $k \geq 1$, we have
$\Iffk[f] \leq (32  s(f))^k \cdot k!$ 
\end{Lem}
\begin{proof}
By Theorem \ref{thm:ik} and Lemma \ref{lem:switching encoding}, we have that
\[
{\frac {\Iffk[f]}{{\ff{n}{k}}}} \leq
\Pr_{\brho \leftarrow \Rkn}[ \deg(f_\brho) =k] \leq
\Pr_{\brho \leftarrow \Rkn}[ \dt(f_\brho) =k] \leq
\frac{(32 s(f))^k}{\binom{n}{k}},
\]
which may be rewritten as the claimed bound.
\end{proof}

%
%

Now we are ready to prove Theorem \ref{thm:main}:

\medskip
\noindent {\bf Theorem \ref{thm:main}.}  \emph{
For any function $f$ and any $\eps > 0$, we have
$
\deg_\eps(f) \leq O(\smax(f) \cdot \log (1/\eps)).
$
}

\begin{proof}
Let $k = \log(1/\eps)$, and 
let $t\ge k$ be some parameter to be determined later. 
We have
$$\sum_{|S| \geq t}\h{f}(S)^2 = \Pr_{\mb{R} \samp \D_f}[|\mb{R}| \geq
  t].$$
Since ${t \choose k}$ is strictly increasing for $t \geq k$, we have
\begin{align*}
\Pr_{\mb{R} \samp \D_f}[|\mb{R}| \geq t] = \Pr_{\mb{R} \samp \D_f}\left[{
    |\mb{R}| \choose k} \geq {t \choose k}\right].
\end{align*}
Now observe that we have
\[{\frac {\Iffk[f]}{k!}} = \Ex_{\mb{R} \samp \D_f}\left[{|\mb{R}| \choose k}\right].\]
Hence
Markov's inequality and Lemma~\ref{lem:Iffk-upper-bound} gives
\begin{align*}
\Pr_{\mb{R} \samp \D_f}[|\mb{R}| \geq t] \leq \frac{\Ex_{\mb{R} \samp \D_f}[{ |\mb{R}| \choose k}]}{{t \choose k}}
\leq \frac{\Iffk[f]/k!}{(t/k)^k} \le \frac{(32  s(f))^k}{(t/k)^k}\;.
\end{align*}
Choosing $t = 64  s(f) \cdot k$ gives 
\[ \Pr_{\mb{R} \samp \D_f}[|\mb{R}| \geq t] \leq \frac{1}{2^k} \le \eps.\] 
Overall, we have
\[
\deg_\eps(f) \le t = \allowbreak O(s(f)\cdot\log(1/\eps)).
\]
\end{proof}

%

We note that the relations between influence moments and Fourier concentration  that are established in \cite[Section 4]{Tal:14tightbounds} can also be used to obtain Theorem \ref{thm:main} from Lemma \ref{lem:Iffk-upper-bound}.  That work \cite[Section 4]{Tal:14tightbounds} also shows that bounded $k$-th influence moments imply  bounded Fourier $L_1$ spectral norm on the $k$-th level, which in turn implies Fourier concentration on a small number of Fourier coefficients (smaller than the trivial ${n \choose k}$ bound on the number of coefficients at degree $k$). These results can be used with Lemma \ref{lem:Iffk-upper-bound} to establish that functions with bounded max sensitivity have sparse Fourier spectra.
%
\begin{Cor}\label{thm:Fourier concentration implications}
	Let $f$ be a Boolean function with sensitivity $s$. For some absolute constant $c$, 
\begin{itemize}
\item $\sum_{S:|S|=k}{|\hat{f}(S)|} \leq (cs)^k$.
\item  $f$ can be $\eps$-approximated in $L_2$ by a polynomial with  at most $s^{c \cdot s \cdot \log(1/\eps)}$ monomials.
\end{itemize}
\end{Cor}
\ignore{\pnote{Moved and condensed}}

\subsection{On the tightness of the Fourier concentration bound}
\label{sec:tightness}

Recall that Conjecture \ref{conj:N} asserts the existance of $c >0$ such that $\deg(f) \leq s(f)^c$, where $c$ needs to be at least $2$. In contrast, we have shown that $\deg_\eps(f) \leq s(f)\log(1/\eps)$. A possible apporach to the Conjecture might be to tradeoff between the exponents of $s$ and $\log(1/\eps)$, by showing bounds of the form $\deg_{\eps}(f)  = O(s(f)^c \cdot \log (1/\eps)^{\gamma})$ for $c \geq 1$ and $\gamma < 1$. We prove two claims about such improvements.

\begin{enumerate}
\item Any bound with constants $(c,\gamma)$ where $\gamma < 1$ will imply Conjecture \ref{conj:1}. 
Concretely, we will show if  $\deg_{\eps}(f)  = O(s(f)^c \cdot \log (1/\eps)^{\gamma})$ for all $\eps>0$ and $\gamma < 1$, then Conjecture \ref{conj:N} holds. 

\item For a bound to hold with constants $(c,\gamma)$ where $\gamma <1$, we would need to have  $c>1$. Concretly, for any positive integer $s$ and any $\eps \ge \Omega(1/s^s)$ there exists a Boolean function with sensitivity at most $s$ and $\deg_{\eps}(f)=\Omega(s \cdot \frac{\log 1/\epsilon}{\log\log 1/\eps})$. 
\end{enumerate}

%

\begin{Lem}
\label{lem:gamma}
If there exist constants $c >0$ and $\gamma < 1$ such that for all $\eps >0$, $\deg_\eps(f) \leq O(s(f)^c\log(1/\eps)^\gamma)$, then $\deg(f) \leq O(s(f)^{c/(1 - \gamma)}$ hence Conjecture \ref{conj:N} holds true.
\end{Lem}
\begin{Proof}
Let $d$ denote $\deg(f)$ and take $\eps = 2^{-3d}$. For $\eps$ so small, we have $\deg_{\eps}(f) = \deg(f) = d$ as any Fourier coefficient is of magnitude at least $2^{-d}$ and the Fourier tail could be either $0$ or at least $2^{-2d}$. Thus, we would get $$d = \deg_{\eps}(f)  \le O(s(f)^{c} \cdot \log(1/\eps)^{\gamma}) \le O(s(f)^{c}  \cdot (3d)^{\gamma})$$ which is equivalent to $d = O(s(f)^{c/(1-\gamma)})$.
\end{Proof}

For the second claim, we give an example showing that our Fourier-concentration result is essentially tight. Our construction is based on the Hamming code. The proof is in Section \ref{sec:ex}.

\begin{Lem}
\label{lem:ham}
	Let $s$ be a positive integer. For every $\epsilon \ge 0.5 \cdot s^{-s}$,
	there exists a Boolean function with sensitivity at most $s$ such that $\deg_{\eps}(f) = \Omega(s \cdot \frac{\log(1/\epsilon)}{\log\log(1/\epsilon)})$.
\end{Lem}

\section{Applications} 
\subsection{Learning Low-Sensitivity Functions}
\label{sec:learn}

In Valiant's PAC learning model, a learning algorithm tries to learn an unknown concept (function) from a class using a hypothesis function. While hypotheses are easy to model using low-level circuit classes, the question of modeling real-world concept classes accurately is more challenging, and has received a lot of attention in the machine learning literature. A fairly standard assumption is that concepts to be learned satisfy some kind of smoothness: roughly speaking, one often assumes that $x$ and $y$ being sufficiently close {\em generally implies} that $f(x) \approx f(y)$ \cite{BL07,BCV}. This assumption favors local prediction algorithms where the prediction of $f(x)$ is dominated by the values at near neighbors of $x$. While we have not found a completely formal defintion, \cite{BL07} call $f$ smooth {\em when the value of $f$ and its derivative $f'$ at $x$ and $y$ are close whenever $x$ and $y$ are close}. This requires a suitable metric on both the domain and the range of $f$. It is natural to ask what smoothness means for Boolean functions, and how the assumption of smoothness affects the complexity of computing and learning a function $f$.

The work of \cite{GNSTW} proposed low worst-case sensitivity as a notion of smoothness for Boolean functions, motivated by connections to classical analysis; we refer the reader to that paper for more details. We believe that low sensitivity is indeed a realistic assumption for several learning tasks (for instance, the concepts learned in image recognition tasks are arguably not sensitive to most individual pixels).  For the natural definition of the gradient $\nabla f$ of $f$, \cite{GSW:xx} show that low sensitivity of $f$ implies that the gradient also has low sensitivity, meaning that a random neighbor of $x$ is likely to be sensitive to the same set of coordinates as the point $x$ itself. Thus low-sensitivity functions have the property that $f$ and $\nabla f$ are likely to agree at nearby points.

Relevant work along these lines is that of Klivans et al. \cite{KOS:04} on learning noise-stable Boolean functions. One can view low noise-sensitivity under the uniform distribution on inputs as a notion of smoothness which guarantees that nearby points are likely to have similar values. \cite{KOS:04} showed that low (uniform-distribution) noise-sensitivity implies efficient learnability under the uniform distribution, since it implies Fourier concentration. Low sensitivity is a stronger assumption that low noise-sensitivity; \cite[Lemma 15]{GNSTW} shows that it implies a strong {\em pointwise} noise stability bound, which says that for any $x$, most $y$ which are close to it satisfy $f(y)  =f(x)$. In contrast, in the definition of noise-sensitivity  both $x$ and $y$ are chosen randomly. 

Conjecture \ref{conj:N} implies that low-sensitivity functions in fact have strong global structure: they have low polynomial degree and small depth as decision trees, so they can be PAC-learned efficiently under arbitrary distributions in time $O(n^{\poly(s(f))})$.  The best provable bound we are aware of is $n^{\exp(O(s))}$, which follows from Simon's result \cite{Simon82} that any sensitivity-$s$ function depends on at most $\exp(O(s))$ variables.
In this section, we give an efficient algorithm for  learning low-sensitivity functions under the uniform distribution from random examples, and in fact show that exact learning is possible in this model.

We consider the problem of learning a concept $f$ from a class of functions $\calC$ given uniformly random examples labeled according to $f$. 
In the standard PAC learning scenario, the goal is produce a hypothesis $h$ such that with high probability (over the coin tosses of the algorithm and the random examples), $\Pr_{\bx \in \zo^n} [f(\bx) = h(\bx)] \geq 1- \eps$.  We refer to this as learning under the uniform distribution with error $\eps$. In the exact learning problem $\eps$ is taken to be zero; the algorithm is allowed to fail with small probability, but otherwise it must return (an efficiently evaluatable representation of) a hypothesis $h$ that agrees with $f$ on every $x \in \{0,1\}^n$.

The seminal work of \cite{LMN:93} showed that Fourier concentration bounds lead to uniform distribution learning algorithms. Applying their result with Theorem \ref{thm:main} we get
\begin{Thm}
\label{thm:pac}
There is an algorithm that learns the class of sensitivity $s$ functions on $n$ variables under the uniform distribution with error $\eps$ in time $n^{O(s\log(1/\eps))}$.
\end{Thm}
The algorithm runs the low-degree algorithm of \cite{LMN:93} to estimate the Fourier spectrum on the bottom $O(s\log(1/\eps))$ levels. It then returns the sign of the resulting polynomial as its hypothesis. This is already a stronger bound than one obtains by assuming low noise-sensitivity (see \cite[Corollary 17]{KOS:04}), since those results have a polynomial in $1/\eps$ in the exponent of $n$. Further, we can use this to get an exact learning algorithm under the uniform distribution, using the following self-correction result from \cite{GNSTW}.

\begin{Thm}
\cite[Theorem 4]{GNSTW}
\label{thm:sc}
There exists a constant $C$ such that  given  $r:\zo^n \rgta \zo$ such that $\Pr_{x \in \zo^n}[r(x) \neq f(x)] \leq 2^{-C s}$
for some sensitivity $s$ function $f$, there is a self-correction algorithm which when
given an oracle for $r$ and $x \in \zo^n$ as input, queries the oracle
for $r$ at $n^{O(s)}$ points, runs in $n^{O(s)}$ time, and returns
$f(x)$ with probability $0.99$.  
\end{Thm}  

We can use this result to give an exact learning algorithm that outputs a randomized Boolean function as a hypothesis. Let us define precisely what this means. A randomized Boolean function $h$ outputs a Boolean random variable $h(x)$ for each input $x \in \zo^n$, by tossing some random coins and performing a randomized computation. Our learning algorithm uses the examples to output a randomized Boolean function $h$ as a hypothesis. With high probability over the examples, the resulting hypothesis $h$ will have the guarantee that for every $x$, $\Pr_h[h(x) = f(x)] \geq 0.99$ where this probability is only over the coin tosses $h$ uses to evaluate $h(x)$.

\begin{Thm}
\label{thm:exact1}
There is an algorithm for exactly learning the class of sensitivity $s$ functions on $n$ variables under the uniform distribution in time $n^{O(s^2)}$. The output hypothesis $h$ is a randomized Boolean function which requires time $n^{O(s)}$ to evaluate on each input and which satisfies  $\Pr_h[h(x) = f(x)] \geq 0.99$ for every $x$. 
\end{Thm}
\begin{Proof}
We first use Theorem \ref{thm:pac} with $\eps = 2^{-C s}$ where $C$ is the constant from Theorem \ref{thm:sc} to get a learn a hypothesis $r$ such that $\Pr_{x \in \zo^n}[r(x) \neq f(x)] \leq 2^{-C s}$ in time $n^{(O(s^2))}$, and invoke the self-corrector from Theorem \ref{thm:sc} on $r$. Let $h$ denote the randomized Boolean function which is the composition of the self-corrector with $r$. By Theorem \ref{thm:sc}, $h$ has the properties claimed above.
\end{Proof}

The hypothesis above is a randomized Boolean function, which may err with some probability on any input. We can eliminate this error and with high probability learn a hypothesis that is correct on every input, at the price of a small increase in the running time. This requires the following result from \cite{GNSTW}.

\begin{Thm} 
\cite[Theorem 3]{GNSTW}
\label{thm:gnstw}
Let $f:\zo^n \to \pmo$ be a sensitivity $s$ function. Given the values of $f$ on any Hamming ball of radius $2s$ as advice, there is an algorithm which, given $x \in \zo^n$ as input, runs in time $n^{O(s)}$ and computes $f(x)$.
\end{Thm}

\begin{Thm}
\label{thm:exact}
There is an algorithm that exactly learns the class of sensitivity $s$ functions on $n$ variables under the uniform distribution in time $n^{O(s^2\log(n))}$. The output hypothesis is Boolean function which requires time $n^{O(s)}$ to evaluate on each input and which equals $f$ on every input in $\zo^n$ with high probability over the learning examples.
\end{Thm}
\begin{Proof}
We first use Theorem \ref{thm:pac} with $\eps = 1/n^{3s}$. The LMN algorithm runs in time $n^{O(s^2\log(n))}$ and with high probability returns a hypothesis $h$ such that 
\[ \Pr_{\bx \in \zo^n} [f(\bx) \neq h(\bx)] \leq 1/n^{3s}.\] 
We then pick a random ball $\mathcal{B}$ of radius $2s$, and use $h$ to label all the points in it. Since each point in the ball is uniformly random, all the labels will be correct with probability $1 - 1/n^s$. We then use these as advice for the algorithm in Theorem \ref{thm:gnstw}. On an input $x \in \zo^n$, the resulting hypothesis can be evaluated in time $n^{O(s)}$ and return the correct value $f(x)$. 
\end{Proof}

\subsection{The Entropy-Influence Conjecture revisited}
\label{sec:Fourier-Entropy}

In this section, we prove a bound on the entropy of the Fourier spectrum of $f$, 
in terms of the influence of $f$ and its sensitivity.
We begin with the definition of the entropy of the Fourier spectrum of $f$.
\begin{Def}
	The Fourier-entropy of a Boolean function $f:\B^n \to \pmo$ is defined to be
	\[ \ent[f] \eqdef \sum_{S \subseteq [n]} \hat{f}(S)^2 \cdot \log_2\left({\frac1 {\hat{f}(S)^2}}\right)\]
\end{Def}

The Fourier Entropy Influence Conjecture by Kalai and Friedgut  \cite{FrKa:96} states that for any Boolean function $f$,
\[ \ent[f] = O(\I[f]).\]  
We upper bound the Fourier-entropy $\ent[f]$ as a function of the influence and the sensitivity of $f$. 
\begin{Thm}
\label{thm:ei}
For any Boolean function $f:\B^n \to \pmo$, 
\[ \ent[f] \le \I[f] \cdot (2\log s[f]+O(1)).\]
\end{Thm}
This improves on the bound $\ent[f] \le \I[f] \cdot O(\log n+ 1)$ given by O'Donnell, Wright and Zhou \cite{ODonnellWZ11} (who also deduced better bounds for symmetric and block-symmetric functions).

In the remainder of this section, we shall denote by $\W^{k}[f]$ the sum $\sum_{S:|S|=k} {\hat{f}(S)^2}$.
We  use the following Theorem of O'Donnell et al. \cite{ODonnellWZ11}.
\begin{Thm}[\protect{\cite[Theorem~5]{ODonnellWZ11}}]
\label{thm:ODonnell}
	Let $f: \B^n \to \pmo$ be a Boolean function. Then $\sum_{k=0}^{n} \W^{k}[f] \cdot \log \frac{1}{\W^{k}[f]} \le 3 \cdot \I[f]$.
\end{Thm}

We use the following upper bound on the entropy of a distribution. To be slightly more general we state the inequality for all sequences of non-negative numbers and not just for those that sum up to $1$.

\begin{Lem}\label{lemma:entropy_vs_L1}
	Let $(p_1, \ldots, p_m)$ be a sequence of non-negative numbers and let $p = \sum_{i=1}^{m}{p_i}$. Then, $$\sum_{i=1}^{m} p_i \log (1/p_i) \le 2p \cdot \log\left(\sum_{i=1}^{m}{\sqrt{p_i}}\right) + 2p\log(1/p).$$
\end{Lem}
\begin{Proof}
	\begin{align*}
	\sum_{i=1}^{m}{p_i \cdot \log \frac{1}{p_i}} 
	= 2 \cdot \sum_{i=1}^{m}{p_i \cdot \log \frac{1}{\sqrt{p_i}}}	 
	= 2 p \cdot \sum_{i=1}^{m}{\frac{p_i}{p} \cdot \log \frac{1}{\sqrt{p_i}}}
	\le 2p \cdot \log\left( \sum_{i=1}^{m}{\sqrt{p_i}/p}\right)
	\end{align*}
	where in the last inequality we applied Jensen's inequality relying on the concavity of the $\log$ function and the fact that $p_i/p$ is a probability distribution.
\end{Proof}

\begin{proof}[Proof of Theorem \ref{thm:ei}]
For each $k=0, \ldots, n$, we denote the contribution from sets of size $k$ to $\ent[f]$ by 
\[ \ent_k[f] \triangleq \sum_{S:|S|=k}{\hat{f}(S)^2 \cdot \log(1/\hat{f}(S)^2)}.\] 
We apply Lemma~\ref{lemma:entropy_vs_L1} on the sequence of numbers $(\hat{f}(S)^2)_{S:|S|=k}$ to get
\begin{align*}
\ent_k[f] &= \sum_{S:|S|=k}{\hat{f}(S)^2 \cdot \log(1/\hat{f}(S)^2)}	\\
&\le 2\cdot \W^{k}[f]\cdot \log\left(\sum_{S:|S|=k}{|\hat{f}(S)|}\right) \;+\; 2\cdot \W^{k}[f] \cdot \log(1/\W^k[f]).
\end{align*}
We invoke the bound from Theorem~\ref{thm:Fourier concentration implications},  $\sum_{S:|S|=k}{|\hat{f}(S)|} \le (Cs)^{k}$ for some universal constant $C>0$, to get
\[
\ent_k[f] \le  2 \cdot \W^{k}[f]  \cdot \log( (Cs)^k ) \;+\; 2 \cdot \W^{k}[f] \cdot \log \frac{1}{\W^{k}[f]}\;.
\]
Summing $\ent_k[f]$ over $k=0,1, \ldots, n$ we get
\[
\ent[f] \le \sum_{k=0}^{n}{\ent_k[f]} \le 2 \cdot \log(Cs) \cdot \sum_{k=0}^{n} \W^{k}[f] \cdot k \;+\;2\cdot \sum_{k=0}^{n}{\W^{k}[f] \cdot \log \frac{1}{\W^{k}[f]}}\;.
\]
Using  the equality $\sum_{k}{\W^{k}[f] \cdot k} = \I[f]$ and the bound from Theorem~\ref{thm:ODonnell} we get 
\[\ent[f] \le 2 \cdot \log(Cs) \cdot \I[f] + 2 \cdot 3 \cdot \I[f] = \I[f] \cdot ( 2\log s + O(1))\;.\]
\end{proof}

\subsection{The switching lemma for $\dnf$s via moments?}
\label{sec:sl-moments}

Our last application concerns the class of width-$w$ $\dnf$ formulas. In Section \ref{sec:moments} we showed how the switching lemma implies sensitivity moment bounds for $\dnf$s (and $\acz$). Here we show the converse, how a version of the switching lemma can be derived using sensitivity moment bounds. We start by showing that moment bounds for $\dnf$s can be derived from the Satisfiability Coding Lemma of \cite{PPZ97}, who give the following tail bound for the sensitivity:

\begin{Lem}\label{lem:sc}\cite{PPZ97}
Let $f:\zo^n \to \pmo$ be computable by a width-$w$ $\dnf$ formula. Then for $s > w$,
\[ \Pr_{\bx \leftarrow \zo^n}[ s(f,\bx) \geq s] \leq 2^{-s/w}.\]
\end{Lem}

More precisely, their Satifiability Coding Lemma \cite[Lemma 2]{PPZ97} uses the width-$w$ $\dnf$ to construct a randomized encoding where a satisfying input $x$ has  (expected) description length bounded by $(n - s(f,x)/w)$, which then implies the above tail bound (see \cite[Fact 3, Lemma 4]{PPZ97}).  Lemma \ref{lem:dnf-moments} uses this tail bound to derive a moment bound.  Lemma \ref{ex:3} in Section \ref{sec:ex} shows that this bound is tight up to the constant $c$.

\begin{Lem}
\label{lem:dnf-moments}
There exists a constant $c$ such that for every integer $k \geq 1$ and every $f:\zo^n \to \pmo$ that is 
computable by a width-$w$ $\dnf$ formula, we have
 $s^k(f) \leq (ckw)^k$. 
\end{Lem}
\begin{Proof}
We have
\begin{align*}
s^k(f) & = \sum_{s = 1}^n s^k\Pr_{\bx}[s(f,\bx) = s]\\
& \leq \sum_{\ell=1}^{n/w} (\ell w)^k\Pr_{\bx}[s(f,\bx) \in\{(\ell -1)w, \ldots, \ell w -1\}]\\
& \leq \sum_{\ell=1}^{n/w} (\ell w)^k 2^{- (\ell -1)} \\
& \leq (ckw)^k
\end{align*}
for some constant $c$, where we used Lemma \ref{lem:sc} to bound $\Pr_{\bx}[s(f,\bx) \geq (\ell -1)w]$. 
\end{Proof}

If Conjecture \ref{conj:1} holds, plugging these bounds into Corollary \ref{cor:sf+conj1} gives that
there exists $c' >0$ such that for any width $w$ $\dnf$ $f$, 
\begin{align*} 
\Pr_{\brho \leftarrow \Rkn}[ \dt(f_\brho) \geq k]   \leq \frac{ 8^k s^k(f)}{{n \choose k}} \leq \frac{(c'kw)^k}{{n \choose k}}.
\end{align*}
Up to a $k!$ term, this matches the bound from H\aa stad's switching lemma \cite[Lemma 1]{Beame94} which shows
\begin{align*} 
\Pr_{\brho \leftarrow \Rkn}[ \dt(f_\brho) \geq k]   \leq \frac{(7kw)^k}{n^k}.
\end{align*}
Thus proving Conjecture \ref{conj:1} would give  a combinatorial proof of the switching lemma for $\dnf$s which seems very different from the known proofs of H\aa stad \cite{Hastad:86} and Razborov \cite{Raz95}.

\section{Examples}
\label{sec:ex}

\begin{Lem}
\label{lem:gap}
Let $n=2^k-1$. There exists $f:\zo^n \rgta \pmo$ for which
$\dt(f) = \log(n+1)$ whereas $ts(f) = n$.
\end{Lem}
\begin{Proof}
Take a complete binary tree with $n$ internal nodes and $n+1$
leaves. The leaves are alternately labelled $1$ and $-1$ from left to
right, while the
internal nodes are labelled with $x_1,\ldots, x_n$ according to an in-order
traversal of the tree.
The bound on decision tree depth follows from the definition of
$f$. To lower bound $\ts(f)$, we start at the $-1^n$ input and start
flipping bits from $-1$ to $1$ in the order $x_1,\ldots,x_n$. It can be
verified that every bit flip changes the value of the function.
\end{Proof}

\begin{Lem}
\label{lem:ex2}
There exists a Boolean function $g$ on $n$ variables such that any proper walk for $g$ has length $\Omega(n^2)$.
\end{Lem}
\begin{Proof}
Assume that $n$ is a power of $2$ and fix a Hadamard code of length $n/2$.
We define an $n$-variable function $g$ over variables
$x_1,\ldots,x_{n/2}$ and $y_1,\ldots, y_{n/2}$ as follows:
if the string $x_1,\dots,x_{n/2}$ equals the $i$-th codeword in the
Hadamard code of length $n/2$, then the output is $y_i$, otherwise
the output is 0. Note that for any $i \neq j$, if $n$-bit inputs $a,b$ are 
sensitive to $y_i$, $y_j$ respectively then the Hamming distance
between $a$ and $b$ must be at least $n/4$. 
Thus any proper walk must flip at least $n/4$
bits between any two vertices that are sensitive to different
$y_i$s, so the minimum length of any proper walk must be 
at least $n^2/8$.
\end{Proof}

The next example, which shows the tightness of Lemma \ref{lem:dnf-moments} up to the constant $c$ is from \cite{PPZ97} where it is used to show the tightness of Lemma \ref{lem:sc}.
\begin{Lem}
\label{ex:3}
For every $k$, there exists a Boolean function $h$ which is computable by a width-$w$ $\dnf$, such that $\s^k(h) \geq (kw/2)^k$.
\end{Lem}
\begin{Proof}
Let $x \in \zo^{k \times w}$ be a $k \times w$ array of bits. Define the function $h$ to be one if some row contains an odd number of $1$s and $0$ otherwise. Formally, let 
\[ h(x) = \vee_{i=1}^k\oplus_{j=1}^wx_{ij}.\]
To lower bound its moment, note that with probability $2^{-k}$, every row contains an even number of $1$s. For such $x$, the $s(h,x) = kw$. This shows that
$s^k(h) \geq (kw/2)^k$.
\end{Proof}

\subsection*{On the tightness of the Fourier tail bound}

We will construct a low-sensitivity function showing that the Fourier tail bounds in Theorem \ref{thm:main} are nearly tight, completing the proof of Lemma \ref{lem:ham}.

For any positive integer $r$, the Hamming code is a linear code of length $m=2^{r}-1$, with $2^{m-r} = 2^{m}/(m+1)$ codewords, such that any two codewords are of distance at least $3$ from one another.
Take $\Ham_m$ to be the indicator function of the Hamming code of length $m$.
We have that the maximal sensitivity on $0$-inputs for $\Ham_m$ (denoted $s_0(\Ham_m)$) equals $1$, as any non-codeword may be adjacent to at most one codeword (otherwise the minimal distance of the code would be $2$). The maximal
sensitivity on $1$-inputs for $\Ham_m$ (denoted $s_1(\Ham_m)$) equals $m$, as any codeword is adjacent to $m$ non-codewords in the $\{0,1\}^m$.

The Hamming code is a linear subspace $U\subseteq {\mathbb{F}}_2^m$ defined by $r$ linear equations. The dual-code (i.e., the dual subspace $U^{\perp}$) is the Hadamard code. There are $m = 2^{r}-1$ non-zero vectors in $U^{\perp}$, all  of them with Hamming-weight $(m+1)/2$.
It is easy to check that the Fourier transform of $\Ham_m:\{0,1\}^m \to \{0,1\}$ (note that we are viewing $\Ham_m$ as a function to $\{0,1\}$ and not $\{-1,1\}$) is 
\[ \Ham_m(x) = \sum_{\alpha \in U^{\perp}} {\frac{1}{m+1} \cdot (-1)^{\sum_{j=1}^{m} \alpha_j x_j}}. \]

For any integer $\ell$, take $f = \OR_{m} \circ \Ham_{m} \circ \Parity_{\ell}$. To be more precise, the function is defined over $m^2 \cdot \ell$ variables $(x_{i,j,k})_{i\in[m],j\in [m], k \in [\ell]}$ as follows:
\[
f(x) = \bigvee_{i=1}^{m} \Ham_{m}( y_{i,1}, \ldots, y_{i,m}),\quad\text{where}\quad
y_{i,j} = \Parity_{\ell}(x_{i,j,1}, \ldots, x_{i,j,\ell})\;.
\]

\begin{Lem}\label{lem:sens f}
The sensitivity of $f$ is  $m \cdot \ell$.	
\end{Lem}
\begin{proof}
	The sensitivity of $g = \OR_m \circ \Ham_m$ is at most $m$ since:
	(1) the sensitivity of $1$-inputs of $g$ is at most the sensitivity of $1$-inputs of $\Ham_m$, and (2) the sensitivity of $0$-inputs of $g$ is at most the sensitivity of $0$-inputs of $\Ham_m$ times $m$.
	To deduce an upper bound on the sensitivity of $f$ we use the fact that for any two functions $f_1, f_2$ we have $s(f_1 \circ f_2) \le s(f_1) \cdot s(f_2)$.
	This gives $s(f) \le s(g) \cdot s(\Parity_\ell) \le m \cdot \ell$.
	
	To show that the sensitivity is at least $m \cdot \ell$ take $x_{i,j,k}$ such that for exactly one $i$, the vector $(y_{i,1},\ldots, y_{i,m})$ is a codeword of the Hamming-code.
	The value of $f$ on $x$ is $1$, and changing any one of the bits $\{x_{i,j,k}: j \in [m], k\in [\ell]\}$ yields an input $x'$ such that $f(x') = 0$. This shows that the sensitivity of $f$ on $x$ is at least $m \cdot \ell$, completing the proof. 
\end{proof}

\begin{Lem}\label{lem:Fourier tails f}
	$f$ has Fourier weight at least $0.5 \cdot m^{-m}$ on the $(m \cdot \frac{m+1}2 \cdot \ell)$-th  layer.
\end{Lem}
\begin{proof}
We view $f$ as a function $f:\{0,1\}^{m^2\ell} \to \{0,1\}$ and write it as a multilinear polynomial in its inputs. Since the Fourier transform is the unique multilinear polynomial equaling $f$ on $\{0,1\}^{m^2\ell}$, we can ``read''  the Fourier coefficients of $f$ from the polynomial.
Then we translate this to the Fourier coefficients of $f':\{0,1\}^{m^2\ell} \to \{-1,1\}$ defined by $f'(x) = (-1)^{f(x)} = 1-2f(x)$.

We write a polynomial agreeing with $\OR_m$ on $\{0,1\}^m$:
\[
\OR_m(z_1, \ldots, z_m) = 1 - \prod_{i=1}^{m}(1-z_i)
= \sum_{\emptyset \neq S \subseteq [m]} (-1)^{|S|} \prod_{i\in S}{z_i}.
\]
We substitute $\Ham_m(y_{i,1},\ldots, y_{i,m})$ for each $z_i$ and then substitute $(\sum_{k=1}^{\ell}{x_{i,j,k}}\mod 2)$ for each $y_{i,j}$ to get
\begin{align*}
f(x) &= \sum_{\emptyset \neq S \subseteq [m]} 
 (-1)^{|S|} \cdot \prod_{i \in S} \Ham_m(y_{i,1}, \ldots, y_{i,m})\\
 &= \sum_{\emptyset \neq S \subseteq [m]} 
 (-1)^{|S|} \cdot \prod_{i \in S} \left(\sum_{\alpha_i \in U^{\perp}} \frac{1}{m+1} \cdot (-1)^{\sum_{j=1}^{m}{\alpha_{i,j} \cdot y_{i,j}}} \right)\\
 &= \sum_{\emptyset \neq S \subseteq [m]} 
 (-1)^{|S|} \cdot \frac{1}{(m+1)^{|S|}}\cdot \sum_{(\alpha_i)_{i\in S}} (-1)^{\sum_{i\in S}\sum_{j=1}^{m}{\alpha_{i,j} \cdot y_{i,j}}}\\
 &= \sum_{\emptyset \neq S \subseteq [m]} 
 (-1)^{|S|} \cdot \frac{1}{(m+1)^{|S|}}\cdot \sum_{(\alpha_i)_{i\in S}} (-1)^{\sum_{i\in S}\sum_{j=1}^{m}\sum_{k=1}^{\ell} {\alpha_{i,j} \cdot x_{i,j,k}}}.
\end{align*}
Gathering the terms according to the characters gives the Fourier expansion of $f$.
For $S = [m]$ and a fixed set of nonzero vectors $\alpha_1, \ldots, \alpha_{m}$  in $U^{\perp}$, the term 
$$
 (-1)^{m} \cdot \frac{1}{(m+1)^{m}} \cdot (-1)^{\sum_{i=1}^{m}\sum_{j=1}^{m}\sum_{k=1}^{\ell} {\alpha_{i,j} \cdot x_{i,j,k}}}
$$
appears in the Fourier expansion of $f$, since it is the only term that appears that is a constant times the character 
\[ (-1)^{\sum_{i=1}^{m}\sum_{j=1}^{m}\sum_{k=1}^{\ell} {\alpha_{i,j} \cdot x_{i,j,k}}}.\]
In other words, the Fourier coefficient of $f$ corresponding to the set $\{(i,j,k) :\alpha_{i,j} = 1\}$ equals $(-1)^{m}/(m+1)^m$. Furthermore, this set is of size $\ell \cdot m \cdot (m+1)/2$, since every $\alpha$ is of weight $(m+1)/2$.
It follows that the Fourier-coefficient of $f' = 1-2f$ corresponding to the same set equals 
\[ \frac{-2\cdot (-1)^m}{(m+1)^m}.\]
As there are $m^m$ choices for non-zero vectors $\alpha_1, \ldots, \alpha_m$ in $U^{\perp}$, the Fourier weight at level  $\ell \cdot m \cdot (m+1)/2$ of $f'$ is at least 
\[ m^m \cdot \frac{4}{(m+1)^{2m}} \geq 0.5 \cdot m^{-m}\] 
for any positive integer $m$.
\end{proof}

\begin{proof}[Proof of Lemma \ref{lem:ham}]
	Choose the maximal integer $m$ such that  $0.5 \cdot m^{-m} \ge \epsilon$. This gives $m = \Theta(\frac{\log(1/\epsilon)}{\log\log(1/\epsilon)})$.
	By assumption on $\epsilon$, $m \le s$. Choose $\ell = \lfloor{s/m\rfloor}$ and take $f = \OR_{m} \circ \Ham_m \circ \Parity_{\ell}$ as above.
	Then, by Lemma~\ref{lem:sens f} the sensitivity of $f$ is at most $m\cdot \ell \le s$.
	By Lemma~\ref{lem:Fourier tails f} the weight at level 
\[ m \cdot \frac{m+1}{2} \cdot \ell = \Theta\left(s \cdot \frac{\log(1/\epsilon)}{\log\log(1/\epsilon)}\right)\] 
is at least $\epsilon$.
\end{proof}

\section{Open questions}
\label{sec:open}

We hope that this work will stimulate further research on the sensitivity graph $G_f$ and on complexity measures associated with it.  
In this context, we'd like to highlight Conjecture \ref{conj:1} which we feel is of independent interest, and if true, implies the robust sensitivity conjecture (Conjecture \ref{conj:moments}) by Corollary \ref{cor:sf+conj2}.  

Another natural question is whether the reverse direction of the robust sensitivity conjecture also holds: for every $k$, does there exist $a'_k, b'_k$ such that $\E[\bolds^k] \leq a'_k\E[\bd^k] +b'_k$? Can one relate this question to a statement about graph-theoretic (or other) complexity measures?

\eat{Returning to the notion of tree sensitivity,
one interesting direction is to bound the diameter of the
sensitive trees that we allow. Let us define $\ts_\newgreen{r}(f)$ to be the maximum tree
sensitivity of $f$ when we restrict to sensitive trees of radius at most 
$\newgreen{r}$.

\begin{Ques}
Does there exist some constant $\newgreen{r}$ such that for all $n$ and all functions
$f:\zo^n \rgta \pmo$, $\ts_\newgreen{r}(f) \geq \dt(f)$?
\end{Ques}

Since $\ts_\newgreen{r}(f) \leq \smax(f)^{O(\newgreen{r})}$, an affirmative
answer to this question would resolve Conjecture \ref{conj:N}.
We note that the stronger statement that $\ts_\newgreen{r}(f) \geq \ts(f)$ for some constant $d$ is not true.
For the read-once decision tree from Lemma \ref{lem:gap}, since
$s(f) \leq \log(n)$, any \newgreen{radius-$r$} sensitive tree has at most
$O(\log^\newgreen{r}(n))$ sensitive edges. But since $\ts(f) \geq n$, we require $\newgreen{r} = \Omega(\log(n)/\log\log(n))$ to
get full tree sensitivity.}

The graph $G_f$ consists of a number of connected components.  This component structure naturally
suggests another complexity measure:

\begin{Def} \label{def:cdim}
For $x \in \{0,1\}^n$, the \emph{component dimension of $f$ at $x$},
denoted $\cdim(f,x)$, is the dimension of the connected component of $G_f$ that contains $x$
(i.e. the number of coordinates $i$ such that $x$'s component contains at least one edge in the $i$-th direction).  
We define $\cdim(f)$ to be $\max_{x \in \{0,1\}^n} \cdim(f,x)$.
\end{Def}
It is easy to see that
$\cdim(f) \geq \ts(f) \geq s(f),$
and thus a consequence of Conjecture \ref{conj:1} is that $\cdim(f) \geq
\dt(f)$; however we have not been able to prove a better lower bound for $\cdim(f)$ in terms
of $\dt(f)$ than that implied by Theorem \ref{thm:stree-dtree}. We
note that $\cdim(f)$ and $\ts(f)$ are not polynomially related, since  the
addressing function shows that the gap between them can be exponential.

The problem of PAC-learning low-sensitivity functions under arbitrary distributions is an intriguing open problem.
Conjecture \ref{conj:N} implies that it should be possible to PAC learn the class of sensitivity-$s$ functions in time $n^{\poly(s)}$ under arbitrary distributions.  While we have shown this holds true under the uniform distribution, we do not know how to do better than $n^{\exp(O(s))}$ for arbitrary distributions.

\subsection*{Acknowledgments}
We thank Yuval Peres and Laci Lovasz for pointing us to Sidorenko's theorem and the related literature. We also thank Yuval for useful discussions, and thank David Levin and Yuval Peres for letting us present the proof of Lemma \ref{lem:peres} here. We thank Gagan Aggarwal for showing us a combinatorial proof of Sidorenko's theorem.  We thank D. Sivakumar for discussions about Section \ref{sec:learn}, and for drawing our attention to relevant work in machine learning.


\bibliography{GSTW-bib}





\end{document}